\documentclass[epj,numbook]{svjour}
\usepackage{epsfig,rotate}
\begin{document}
\title{Non-equilibrium critical behavior of O(n)-symmetric systems} 

\subtitle{Effect of reversible mode-coupling terms and dynamical anisotropy}

\dedication{Dedicated to Franz Schwabl on the occasion of his 60th birthday.}

\author{Uwe C. T\"auber\inst{1} \and Jaime E. Santos\inst{1} \and 
        Zolt\'an R\'acz\inst{2} }

%
%

\institute{Institut f\"ur Theoretische Physik, Technische Universit\"at 
        M\"unchen, James-Franck-Stra\ss e, 85747 Garching, Germany;
        \email{utaeuber@physik.tu-muenchen.de, jesantos@physik.tu-muenchen.de} 
        \and Institute for Theoretical Physics, E\"otv\"os University,
        1088 Budapest, Puskin u. 5--7, Hungary; \email{racz@hal9000.elte.hu},
        and Department of Physics --- Theoretical Physics, 
        University of Oxford, 1 Keble Road, Oxford OX1 3NP, U.K.;
        \email{z.racz1@physics.ox.ac.uk} }
\date{\today}
%
\abstract{\rm
Phase transitions in non-equilibrium steady states of $O(n)$-symmetric models 
with reversible mode couplings are studied using dynamic field theory and the
renormalization group.
The systems are driven out of equilibrium by dynamical anisotropy in the noise 
for the conserved quantities, i.e., by constraining their diffusive dynamics to
be at different temperatures $T^\parallel$ and $T^\perp$ in $d_\parallel$- and 
$d_\perp$-dimensional subspaces, respectively. 
In the case of the Sasv\'ari-Schwabl-Sz\'epfalusy (SSS) model for planar ferro-
and isotropic antiferromagnets, we assume a dynamical anisotropy in the noise 
for the non-critical conserved quantities that are dynamically coupled to the 
non-conserved order parameter.
We find the equilibrium fixed point (with isotropic noise) to be stable with 
respect to these non-equilibrium perturbations, and the familiar equilibrium 
exponents therefore describe the asymptotic static and dynamic critical 
behavior.
Novel critical features are only found in extreme limits, where the ratio of 
the effective noise temperatures $T^\parallel / T^\perp$ is either zero or 
infinite.
On the other hand, for model J for isotropic ferromagnets with a conserved 
order parameter, the dynamical noise anisotropy induces effective long-range 
elastic forces, which lead to a softening only of the $d_\perp$-dimensional 
sector in wavevector space with lower noise temperature 
$T^\perp < T^\parallel$. 
The ensuing static and dynamic critical behavior is described by power laws of 
a hitherto unidentified universality class, which, however, is not accessible 
by perturbational means for $d_\parallel \geq 1$.
We obtain formal expressions for the novel critical exponents in a double
expansion about the static and dynamic upper critical dimensions and 
$d_\parallel$, i.e., about the equilibrium theory.
} 
\maketitle

\section{Introduction}
 \label{introd}

The equilibrium properties of a generic system in contact with a heat bath are 
entirely determined by interactions (the dynamics plays a role only in 
providing the necessary mixing in phase space). 
In contrast, dynamics is important in non-equilibrium steady states (NESS) 
where competing dynamics (i.e. contacts with more than one heat baths or 
driving fields) generate fluxes of energy, mass, etc., or, equivalently, yield 
non-zero steady-state probability currents in phase space. 
This difference between equilibrium and NESS has often been illustrated by 
using an electrodynamical analogy \cite{bearev}: {\em equilibrium $\equiv$ 
electrostatics} while {\em NESS $\equiv$ magnetostatics}. 
The above analogy can actually be developed a bit further by noting that 
magnetostatics can be described in terms of interactions between currents 
and thus asking if NESS could also be described in terms of some effective 
interactions, characterizing the stationary probability distribution. 
  
The search for such effective interactions has been going on for some time.
The simplest version of this search consists of taking the logarithm of the 
steady-state distribution function for small-size systems and looking for the 
dominant interaction in the emerging effective hamiltonian. 
Unfortunately, this approach did not turn out to be very useful since, 
generically, all the interactions, of any range, which are consistent with the 
symmetries of the system are generated --- and the lack of significant 
differences in the magnitude of the couplings renders the identification of the
dominant interactions impossible.

A somewhat more sophisticated approach is based on the extention of 
universality concepts to phase transitions in NESS. 
The idea here is that the universality class of a phase transition in NESS 
provides information about the range of interactions generated by the competing
dynamics and, furthermore, also gives the exponent of the long-distance, 
power-law decay in case of long-range interactions. 
This approach has yielded some very interesting results in connection with 
non-equilibrium perturbations imposed on the relaxational models A 
(non-conserved order parameter dynamics) and B (conserved order parameter 
dynamics), using the terminology of Ref.~\cite{hohhal}. 

In particular, phase transitions in systems with model-A type of dynamics have 
been shown to be remarkably robust against the introduction of various 
competing dynamics which are local and do not conserve the order parameter 
\cite{grins1}, including competing dynamics which breaks the discrete symmetry
of the system \cite{kevbea}. 
This robustness of the {\em critical} behavior also persists when the competing
dynamics comes from a reversible mode coupling to a non-critical conserved 
field \cite{uwezol}. 
Thus there is large class of non-equilibrium steady states where the character 
of interactions is not modified by the presence of thermodynamic fluxes or, in 
other words, the probability currents in phase space become irrelevant for the
large-scale behavior. 
Thinking in terms of the electrodynamic analogy, one may say that this 
corresponds to the magnetostatic problem of a steady electric current along a 
straight line, which can be reduced to an electrostatic problem after an 
appropriate coordinate transformation.
 
Another class of competing dynamics is obtained when model-B type dynamics 
competes with external drive \cite{bearev} or with local, {\em anisotropic}, 
order-parameter conserving processes \cite{bearoy,schmit,kevzol}. 
In these cases, long-range interactions do get generated in the NESS and, 
furthermore, a common feature of these interactions is that their angular 
dependence resembles the form of elastic or uniaxial dipolar forces (and for 
some systems even the power-law decay with distance is that occurring for 
dipolar interactions \cite{bearoy,schmit,kevzol}). 
In terms of the electrodynamics analogy, one might say that the fluxes in the 
NESS of these systems are equivalent to loops of electric currents which 
interact, in the first-order multipole expansion, via (pseudo-)dipolar forces.
(However, when drawing this analogy, one should be aware that while successive 
terms in the electrodynamics multipole expansion become more and more 
suppressed, the effective long-range forces appearing in the NESS of models 
with conserved order parameter are very dominant.) 

There are, of course, examples which do not fit easily into this heuristic
straight-line current and loop current classification, which therefore should
not be taken too far; e.g., competition of the usual model A with non-local 
dynamics \cite{drtbra} or with {\em linear} coupling to a conserved field 
\cite{grins2} generates isotropic long-range interactions.
Non-thermally driven steady states occuring, for example, in the prominent 
non-equilibrium universality classes of directed percolation \cite{absorb} and 
Kardar--Parisi--Zhang surface growth \cite{kapezt} probably cannot be described
in terms of such effective interactions at all. 
Furthermore, there are driven systems where the emerging effective interactions
are extremely long-ranged in the sense that the potential is non-integrable
and, consequently, the system does not display thermodynamical behavior in the 
usual sense \cite{mevans,geyink}.

Nevertheless, since a large variety of competing dynamical processes have 
yielded a surprisingly small number of effective universality classes, we feel 
that it is worthwhile to continue the exploration of NESS through studying 
non-equilibrium phase transitions and thus deducing effective interactions.  
The investigation of the stability of the known equilibrium dynamic 
universality classes is important also from another viewpoint. 
Namely, in many experiments probing critical dynamics, it is by no means a 
trivial issue to maintain thermodynamic equilibrium, as relaxation times become
very long close to a second-order phase transition as a consequence of critical
slowing down.
For the interpretation of experimental data, it might therefore be important to
know if the dynamical system is driven to a NESS that is characterized by 
scaling behavior which is distinct from its equilibrium counterpart.

In the present work, we follow up a previous investigation \cite{uwezol} of 
NESS generated in $O(n)$-symmetric systems subject to non-equilibrium 
perturbations.
Here, either a non-conserved order parameter is dynamically coupled to 
non-critical conserved quantities --- this defines the $n$-component 
Sasv\'ari-Schwabl-Sz\'epfalusy (SSS) model \cite{sssmod,sssfth}, which 
incorporates model E for planar ferromagnets (with $n = 2$) \cite{modele} and 
model G for isotropic antiferromagnets ($n = 3$) \cite{modelg} as special 
cases; or the order parameter and the generators of $O(n)$ are identical, which
is realized for $n = 3$ in model J for (idealized) isotropic ferromagnets 
\cite{modelj}.
(For a review on more realistic dynamics for ferromagnetic systems that
includes the effects of dipolar interactions, see Ref.~\cite{erwrev}.)
This field-theoretic renormalization-group (RG) study found that spatially 
{\em isotropic} violation of detailed balance generically leaves the 
equilibrium fixed point stable, thus indicating that the steady-state fluxes
involved here do {\em not} generate any long-range interactions. 
Since {\em dynamical anisotropy} appears to be an essential ingredient for the
generation of pseudo-dipolar effective interactions in NESS, it is natural to 
ask if this was the case for the above $O(n)$-symmetric systems as well. 
More precisely, we allow for dynamical spatially anisotropic noise for the 
conserved quantities.
For the anisotropic non-equilibrium model J, this means that we allow for 
different effective temperatures $T^\parallel > T^\perp$, respectively, 
governing the order parameter noise in $d_\parallel$- and
$d_\perp$-dimensional subspaces ($d_\parallel + d_\perp = d$).
Essentially, this means that there is a non-zero heat current flowing from
the ``hotter subspace'' ($d_\parallel$ dimensions) into the
$d_\perp$-dimensional ``cooler subspace''.
For the anisotropic non-equilibrium SSS model, a similar distinction applies 
for the noise in the Langevin equation that describes the dynamics of the
purely dynamically coupled conserved field, and we may then explore its 
influence, and that of the ensuing effective heat current, on the non-conserved
order parameter dynamics.

In Sec.~\ref{modeqs}, we shall discuss our basic model equations for the
spatially anisotropic non-equilibrium generalizations of Langevin dynamics
appropriate for second-order phase transitions in $O(n)$-symmetric systems.
We start our investigations with the anisotropic non-equilibrium version of the
SSS model in Sec.~\ref{sssmod}, and shall find that for this system of coupled 
Langevin equations with a non-conserved order parameter, the equilibrium 
dynamic scaling fixed point remains stable, and governs the asymptotic critical
behavior. 
Thus, the ensuing static critical exponents are those of the $n$-component 
Heisenberg model, accompanied with the equilibrium SSS strong-scaling dynamic 
exponent $z = d/2$ (at least to one-loop order). 
Novel dynamic and static scaling exponents are found in the extreme situations 
$T^\perp / T^\parallel = 0$ or $\infty$ only, and may be related to the 
results for the isotropic non-equilibrium model studied earlier \cite{uwezol}. 
The latter are of course contained in our present more general study as a
special case.
These results once more underline previous observations that non-equilibrium
generalizations of dynamical models with a {\em non-conserved} order parameter
asymptotically, i.e., in the vicinity of the critical point, display the
scaling behavior of the corresponding equilibrium dynamic universality class.

In Sec.~\ref{modelj}, we consider the anisotropic non-equilibrium model J with 
{\em conserved} order parameter.
In contrast to the anisotropic non-equilibrium SSS model, we shall find that 
only the spatial sector with {\em lower} noise temperature 
$T^\perp < T^\parallel$ becomes soft at the transition, while the
$d_\parallel$-dimensional sector remains uncritical.
As a consequence, long-range elastic (uniaxial pseudo-dipolar for 
$d_\perp = 1$) effective interactions are generated, as in the 
two-tempera\-ture model B \cite{bearoy,schmit,kevzol}. 
The equilibrium dynamic fixed point becomes unstable, and the phase transition
is described by a novel universality class, characterized by reduced upper and 
lower critical dimensions for both statics and dynamics.
For general dimensionality of the soft sector, we find runaway 
renormalization-group flow trajectories, and perturbational methods appear to 
break down.
Presumably, this indicates either strong-coupling scaling behavior, or perhaps
even the absence of a non-equilibrium steady state at the critical point.
However, we can obtain the new exponents in a double expansion about both the 
upper critical dimensions, and the dimensionality of the hard sector 
$d_\parallel$ (which amounts to an expansion about the equilibrium theory).
Yet, as this expansion clearly becomes invalid at some critical value of
$d_\parallel^c < 1$, at least to one-loop order, this formal expansion should 
not be taken too seriously for the description of a real physical system.
At any rate, this model characterized by a {\em conserved} order parameter, and
{\em spatially anisotropic} conserved noise, definitely leads to a novel 
dynamic universality class.
Finally, in Sec.~\ref{sumcon} we summarize our results again, and discuss their
implications.
In the Appendix, we present the Ward identities stemming from the $O(n)$ 
symmetry of the non-equilibrium models under investigation here.

\section{Model equations}
 \label{modeqs}

In this section, we briefly outline the basic model equations for our 
anisotropic, non-equilibrium generalization of both the SSS model \cite{sssmod}
and model J \cite{modelj}.
The equilibrium characteristics of these dynamic models were summarized at 
length in Ref.~\cite{uwezol}, and we refer to this paper and the original 
equilibrium literature (see 
Refs.~\cite{sssmod,sssfth,modele,modelg,modelj,erwrev,dynfun}) for further
details. 
We shall largely use the notations introduced in Ref.~\cite{uwezol}, if not 
explicitly mentioned otherwise.

We consider a second-order phase transition for an $n$-component vector order 
parameter $S_0^\alpha$, $\alpha = 1,\ldots,n$ (we denote unrenormalized 
quantities by the subscript ``$0$'').
As we furthermore assume isotropy in order parameter space, the static critical
properties will be described by an $O(n)$-symmetric $\phi^4$ 
Landau--Ginzburg--Wilson free energy in $d$ space dimensions,
\begin{eqnarray}
        H[\{ S_0^\alpha \}] = \int \! d^dx &&\Biggl\{ {r_0 \over 2} 
        \sum_{\alpha=1}^n S_0^\alpha({\bf x})^2 + {1 \over 2} \sum_{\alpha=1}^n
                \left[ \nabla S_0^\alpha({\bf x} \right)]^2 \nonumber \\
        &&\quad + {u_0 \over 4 !} \left[ \sum_{\alpha=1}^n 
                S_0^\alpha({\bf x})^2 \right]^2 \Biggr\} \ ,
 \label{hamilt}
\end{eqnarray}
where $r_0 = (T - T_c^0) / T_c^0$ is the relative distance from the mean-field 
critical temperature $T_c^0$. 
This effective free energy determines the equilibrium probability distribution
for the vector order parameter $S_0^\alpha$,
\begin{equation}
        P_{\rm eq}[\{ S_0^\alpha \}] = {e^{-H[\{ S_0^\alpha \}] / k_{\rm B} T} 
                \over \int {\cal D}[\{ S_0^\alpha \}] 
                e^{-H[\{ S_0^\alpha \}] / k_{\rm B} T}} \ .
 \label{eqdist}
\end{equation}
Following standard procedures, one may then compute the two independent 
critical exponents, e.g., $\eta$ and $\nu$, by means of perturbation theory 
with respect to the static non-linear coupling $u_0$ and by employing the 
renormalization group procedure, within a systematic expansion in terms of
$\epsilon = 4-d$ about the static upper critical dimension $d_c = 4$.
Here, $\eta$ describes the power-law decay of the order parameter correlation 
function at criticality, $\langle S^\alpha({\bf x}) S^\beta({\bf x}') \rangle 
\propto 1 / |{\bf x} - {\bf x}'|^{d - 2 + \eta}$, or, equivalently, of the 
static susceptibility, $\chi({\bf q}) \propto 1 / q^{2 - \eta}$, and the 
exponent $\nu$ characterizes the divergence of the correlation length as $T_c$ 
is approached, $\xi \propto |T - T_c|^{-\nu}$.
Notice that fluctuations also shift the true transition temperature $T_c$ 
downwards as compared to the mean-field critical temperature $T_c^0$, i.e., 
$r_{0c} = T_c - T_c^0 < 0$.

In order to correctly describe the critical dynamics for an $O(n)$-symmetric
system, one needs to take into account all the slow modes.
Generally, in addition to the order parameter itself, these comprise the 
diffusive modes associated with the conservation law connected with the 
rotational symmetry.
The SSS model thus consists of dynamically coupled Langevin equations for a
{\em non-conserved} $n$-component order parameter $S_0^\alpha$ and 
$n (n-1) / 2$ conserved generalized angular momenta 
$M_0^{\alpha \beta} = - M_0^{\beta \alpha}$ \cite{sssmod}.
Physical realizations of this model are the critical dynamics of the XY model 
($n = 2$), also called model E \cite{hohhal,modele}, with the order parameter 
components $S_0^x$ and $S_0^y$, and the conserved quantity $M_0^{12} = S_0^z$, 
which generates rotations in the $xy$-plane; and the dynamic critical behavior 
of isotropic antiferromagnets ($n = 3$), known as model G \cite{hohhal,modelg},
with $S_0^x$, $S_0^y$, and $S_0^z$ representing the components of the staggered
magnetization, and $M_0^{12} = M_0^z$, $M_0^{23} = M_0^x$, and 
$M_0^{13} = - M_0^y$ denoting the components of the magnetization itself, which
are conserved and can be identified with the generators of the symmetry group
$O(3)$. 

The variables $M_0^{\alpha \beta}$ represent non-critical quantities, and their
coupling to the order parameter fluctuations $S_0^\alpha$ is of purely 
dynamical character. 
Hence it suffices to simply add a quadratic term to the hamiltonian 
(\ref{hamilt}),
\begin{equation}
        H[\{ S_0^\alpha \} , \{ M_0^{\alpha \beta} \}] = 
        H[\{ S_0^\alpha \}] + {1 \over 2} \int \! d^dx \sum_{\alpha > \beta} 
                M_0^{\alpha \beta}({\bf x})^2 \ .
 \label{stasss}
\end{equation}
With this free energy funtional $H$, the coupled non-linear Langevin equations 
defining the SSS model read \cite{sssmod,sssfth,uwezol}
\begin{eqnarray}
        {\partial S_0^\alpha \over \partial t} &=& 
        g_0 \sum_\beta {\delta H \over \delta M_0^{\alpha \beta}} \, S_0^\beta 
        - \lambda_0 {\delta H \over \delta S_0^\alpha} + \zeta^\alpha 
 \label{soplao} \\
        &=& g_0 \sum_{\beta \not= \alpha} M_0^{\alpha \beta} S_0^\beta  
                - \lambda_0 (r_0 - \nabla^2) S_0^\alpha - \nonumber \\
        &&\quad - \lambda_0 {u_0 \over 6} S_0^\alpha 
                \sum_\beta S_0^\beta S_0^\beta + \zeta^\alpha \ ,
 \label{soplan}
\end{eqnarray}
and
\begin{eqnarray}
        {\partial M_0^{\alpha \beta} \over \partial t} &=& 
        - g_0 \left( {\delta H \over \delta S_0^\alpha} \, S_0^\beta 
        - {\delta H \over \delta S_0^\beta} \, S_0^\alpha \right) + \nonumber\\
        &&\qquad + D_0 \nabla^2 {\delta H \over \delta M_0^{\alpha \beta}}
                + \eta^{\alpha \beta}
 \label{samlao} \\
        &=& - g_0 \left( S_0^\alpha \nabla^2 S_0^\beta 
                - S_0^\beta \nabla^2 S_0^\alpha \right) + \nonumber \\
        &&\qquad + D_0 \nabla^2 M_0^{\alpha \beta} + \eta^{\alpha \beta} \ .
 \label{samlan}
\end{eqnarray}
Here, $g_0$ denotes the strength of the reversible, so-called mode-coupling 
terms, and $\zeta^\alpha$ and $\eta^{\alpha \beta}$ represent fluctuating 
forces with zero mean, $\langle \zeta^\alpha({\bf x},t) \rangle = 0$, 
$\langle \eta^{\alpha \beta}({\bf x},t) \rangle = 0$.
In order to fully characterize the dynamics, we furthermore need to specify the
correlations of these stochastic forces; for the order parameter, we simply
assume a Gaussian distribution for the $\zeta^\alpha$ with the second moment
\begin{equation}
        \langle \zeta^\alpha({\bf x},t) \zeta^\beta({\bf x}',t') 
        \rangle = 2 {\widetilde \lambda}_0 \, \delta({\bf x} - {\bf x}') \,
                        \delta( t - t') \, \delta^{\alpha \beta} \ ,
 \label{sopnoi}
\end{equation}
corresponding to non-conserved white noise.
On the other hand, the conservation law for the generalized angular momenta
$M_0^{\alpha \beta}$, and the antisymmetry with respect to the tensor indices
$\alpha,\beta$ implies that $\langle \eta^{\alpha \beta}({\bf x},t) 
\eta^{\gamma \delta}({\bf x}',t') \rangle \propto - \nabla^2 
\delta({\bf x} - {\bf x}') \delta( t - t') \left( \delta^{\alpha \beta} 
\delta^{\gamma \delta} - \delta^{\alpha \delta} \delta^{\beta \gamma} \right)$.
This functional form, prescribed by the Einstein relation ensuring that at long
times the equilibrium distribution (\ref{eqdist}) with the free energy 
(\ref{stasss}) will be attained, provides us with the possibility to impose a 
{\em spatially anisotropic} form of detailed-balance violation through the
prescription
\begin{eqnarray}
        \langle \eta^{\alpha \beta}({\bf x},t) 
                \eta^{\gamma \delta}({\bf x}',t') \rangle &=&
        - 2 \left( {\widetilde D}_0^\parallel \nabla_\parallel^2 + 
                {\widetilde D}_0^\perp \nabla_\perp^2 \right)
                \delta({\bf x} - {\bf x}') \times \nonumber \\
        &&\!\!\!\!\! \times \delta( t - t') \left( \delta^{\alpha \beta} 
        \delta^{\gamma \delta} - \delta^{\alpha \delta} \delta^{\beta \gamma} 
                \right) \ .
 \label{samnoi}
\end{eqnarray}

We may interpret this as follows. First, consider the special case 
${\widetilde D}^\parallel_0 = {\widetilde D}_0^\perp = {\widetilde D}_0$, which
is in fact the model investigated in Ref.~\cite{uwezol}.
We can then identify ${\widetilde \lambda}_0 / \lambda_0 = k_{\rm B} T_S$ and 
${\widetilde D}_0 / D_0 = k_{\rm B} T_M$ as the temperatures of the heat baths 
coupling to the order parameter and conserved quantities, respectively. 
For $T_S = T_M$, obviously detailed balance holds.
More generally, the ratio $T_0 = T_M / T_S$ describes the extent to which this
equilibrium condition is violated (in Ref.~\cite{uwezol}, $\Theta_0 = 1/T_0$ 
was used instead); for $T_0 < 1$, energy flows from the order parameter heat 
bath into the conserved-quantities heat bath, and vice versa for $T_0 > 1$.
Notice that as we are really interested in the vicinity of the critical point,
$T_S \approx T_c$, and $T_0$ thus gives the heat bath temperature of the 
generalized angular momenta measured in terms of $T_c$.
While new critical behavior ensues for either $T_0 = 0$ or $T_0 = \infty$, a 
one-loop renormalization-group analysis shows that for {\em any} 
$0 < T_0 < \infty$ the asymptotic critical properties are those of the 
equilibrium model, i.e., the renormalized $T \to 1$ under scale
transformations \cite{uwezol}.
In the anisotropic generalization (\ref{samnoi}), we furthermore allow for
different temperatures in $d_\parallel$- and $d_\perp$-dimensional sectors in
space (with $d_\parallel + d_\perp = d$), with associated distinct temperatures
(again, essentially measured with respect to $T_S \approx T_c$),
\begin{equation}
        T_0^{\parallel / \perp} = {{\widetilde D}_0^{\parallel / \perp} \over
                D_0} \, {\lambda_0 \over {\widetilde \lambda}_0} \ .
 \label{snoitm}
\end{equation}
Therefore there appears an additional degree of freedom here, namely the ratio
\begin{equation}
        \sigma_0 = T_0^\perp / T_0^\parallel
 \label{snoitr}
\end{equation}
of the conserved-noise heat bath temperatures in the transverse and parallel
sectors, respectively, which without loss of generality we may assume to be in
the interval $0 \leq \sigma_0 \leq 1$, where $\sigma_0 = 1$ obviously
corresponds to the equilibrium situation.
The one-loop renormalization-group flow equations for this non-equilibrium SSS 
model with spatially anisotropic conserved noise will be derived and studied in
Sec.~\ref{sssmod}.

Model J (in the nomenclature coined in Ref.~\cite{hohhal}) corresponds to the
situation where the conserved quantities associated with the $O(n)$ symmetry
are in fact identical to the order parameter fluctuations themselves.
The physical realization, with $n = 3$, is the critical dynamics of isotropic
ferromagnets, with $S_0^x$, $S_0^y$, and $S_0^z$ denoting the three components
of the conserved magnetization vector \cite{modelj,erwrev}.
The three coupled Langevin equations for model J read
\cite{modelj,erwrev,uwezol}
\begin{eqnarray}
        {\partial S_0^\alpha \over \partial t} &=& 
                - g_0 \sum_{\beta,\gamma} \epsilon^{\alpha \beta \gamma} 
                {\delta H \over \delta S_0^\beta} \, S_0^\gamma 
        + \lambda_0 \nabla^2 {\delta H \over \delta S_0^\alpha} + \zeta^\alpha 
 \label{ferlao} \\
        &=& - g_0 \sum_{\beta,\gamma} \epsilon^{\alpha \beta \gamma} 
                S_0^\beta \nabla^2 S_0^\gamma 
                + \lambda_0 \nabla^2 (r_0 - \nabla^2) S_0^\alpha + \nonumber \\
        &&\quad + \lambda_0 {u_0 \over 6} \nabla^2 S_0^\alpha 
                \sum_\beta S_0^\beta S_0^\beta + \zeta^\alpha \ ,
 \label{ferlan}
\end{eqnarray}
where $g_0$ again denotes the reversible mode-coupling constant, and 
$\langle \zeta^\alpha({\bf x},t) \rangle = 0$ for the stochastic forces.
As now the order parameter is {\em conserved}, we may introduce spatially
anisotropic noise in the correlator of the associated conserved noise, in 
analogy with Eq.~(\ref{samnoi}) for the generalized angular momenta of the SSS 
model,
\begin{eqnarray}
        \langle \zeta^\alpha({\bf x},t) \zeta^\beta({\bf x}',t') \rangle &=& 
        - 2 \left( {\widetilde \lambda}_0^\parallel \nabla_\parallel^2 
                + {\widetilde \lambda}_0^\perp \nabla_\perp^2 \right)
                \delta({\bf x} - {\bf x}') \times \nonumber \\
        &&\quad \times \delta( t - t') \, \delta^{\alpha \beta} \ .
 \label{fernoi}
\end{eqnarray}
In the corresponding isotropic variant where 
${\widetilde \lambda}_0^\parallel = {\widetilde \lambda}_0^\perp$, a simple
rescaling of the non-linear couplings $u_0$ and $g_0$ can absorb the effects of
detailed-balance violation entirely; this demonstrates immediately that this 
model is asymptotically governed by the equilibrium critical exponents 
\cite{uwezol}.
Here, on the other hand, we may define different effective temperatures for the
longitudinal and transverse sectors, respectively,
\begin{equation}
        T_0^{\parallel / \perp} = 
        {\widetilde \lambda}_0^{\parallel / \perp} / \lambda_0 \ ,
 \label{jnoitm}
\end{equation}
and thus once again a novel variable emerges, namely the ratio 
$\sigma_0 = {\widetilde \lambda}_0^\perp / {\widetilde \lambda}_0^\parallel = 
T_0^\perp / T_0^\parallel$, with $0 < \sigma_0 < 1$.
The dramatic implications of this spatially anisotropic conserved noise will be
investigated in Sec.~\ref{modelj}.

\section{The anisotropic non-equilibrium SSS model}
 \label{sssmod}

We now proceed by considering the critical behavior of our anisotropic 
non-equilibrium version of the SSS model \cite{sssmod}, as defined by the 
Langevin equations (\ref{soplan}) and (\ref{samlan}), with the noise 
correlators (\ref{sopnoi}) and (\ref{samnoi}), respectively. 
In Sec.~\ref{sssren}, we perform the perturbational renormalization to one-loop
order, following the procedures that were already employed in 
Ref.~\cite{uwezol}.
These in turn constitute the appropriate generalization of the equilibrium
renormalization scheme, see Refs.~\cite{sssfth,oerjan}.
From the renormalization constants ($Z$ factors) that render the field theory
finite in the ultraviolet (UV), one may then derive the RG flow functions which
enter the Gell-Mann--Low equation.
This partial differential equation describes how correlation functions change 
under scale transformations.
In the vicinity of an RG fixed point, the theory becomes scale-invariant, and
the information previously gained about the UV behavior can thus be employed to
access the physically interesting power laws governing the infrared (IR) regime
at the critical point ($\tau \propto T - T_c \to 0$) for long wavelengths 
(wavevector ${\bf q} \to 0$) and low frequencies ($\omega \to 0$).  

\subsection{Perturbation theory and renormalization}
 \label{sssren}

\subsubsection{Dynamic field theory}

As a first step, we translate the Langevin equations (\ref{soplan}) and 
(\ref{samlan}), with (\ref{sopnoi}) and (\ref{samnoi}), into a dynamic field
theory, following standard procedures \cite{dynfun,sssfth,uwezol}.
This results in a probability distribution for the dynamic fields $S_0^\alpha$
and $M_0^{\alpha \beta}$,
\begin{eqnarray}
        P[\{ S_0^\alpha , M_0^{\alpha \beta} \}] &\propto&
        \int {\cal D}[\{ i {\widetilde S}_0^\alpha \}] \int 
        {\cal D}[\{ i {\widetilde M}_0^{\alpha \beta} \}] \times \nonumber \\
        &&\times \, e^{J[\{ {\widetilde S}_0^\alpha \}, \{ S_0^\alpha \},
        \{ {\widetilde M}_0^{\alpha \beta} \}, \{ M_0^{\alpha \beta} \}]} \ ,
 \label{genfun}
\end{eqnarray}
with the statistical weight given by Janssen-De~Dominicis functional
$J = J_{\rm har} + J_{\rm rel} + J_{\rm mc}$.
Its harmonic part, in terms of the original dynamic fields 
$S_0^\alpha({\bf x},t)$, $M_0^{\alpha \beta}({\bf x},t)$ and the 
auxiliary fields ${\widetilde S}_0^\alpha({\bf x},t)$, 
${\widetilde M}_0^{\alpha \beta}({\bf x},t)$ reads
\begin{eqnarray}
        &&J_{\rm har}[\{ {\widetilde S}_0^\alpha \} , \{ S_0^\alpha \} ,
        \{ {\widetilde M}_0^{\alpha\beta} \} , \{ M_0^{\alpha\beta} \}] = 
                                                        \nonumber \\
        &&= \int \! d^dx \int \! dt \Biggl\{ \sum_\alpha {\widetilde \lambda}_0
        \, {\widetilde S}_0^\alpha {\widetilde S}_0^\alpha - \nonumber \\
        &&\qquad \qquad - \sum_\alpha {\widetilde S}_0^\alpha \left[ 
        {\partial \over \partial t} + \lambda_0 \left( r_0 - \nabla^2 \right) 
                \right] S_0^\alpha - \nonumber \\ 
        &&\qquad \qquad - \sum_{\alpha > \beta} {\widetilde M}_0^{\alpha \beta}
        \left( {\widetilde D}_0^\parallel \nabla_\parallel^2 + 
                {\widetilde D}_0^\perp \nabla_\perp^2 \right) 
                        {\widetilde M}_0^{\alpha \beta} - \nonumber \\
        &&\qquad \qquad - \sum_{\alpha > \beta} {\widetilde M}_0^{\alpha \beta}
        \left( {\partial \over \partial t} - D_0 \nabla^2 \right) 
                M_0^{\alpha \beta} \Biggr\} \ , 
 \label{harfun}
\end{eqnarray}
while the static non-linearity leads to a relaxation vertex
\begin{equation}
        J_{\rm rel}[\{ {\widetilde S}_0^\alpha \} , \{ S_0^\alpha \}] = 
        - \lambda_0 {u_0 \over 6} \int \! d^dx \int \! dt \sum_{\alpha,\beta} 
        {\widetilde S}_0^\alpha S_0^\alpha S_0^\beta S_0^\beta \ ,
 \label{relfun}
\end{equation}
and the purely dynamic couplings generate the mode-coupling vertices
\begin{eqnarray}
        &&J_{\rm mc}[\{ {\widetilde S}_0^\alpha \} , \{ S_0^\alpha \} ,
        \{ {\widetilde M}_0^{\alpha \beta} \} , \{ M_0^{\alpha \beta} \}] = 
                                                        \nonumber \\
        &&= g_0 \int \! d^dx \int \! dt \sum_{\alpha , \beta} \Biggl\{ 
        {\widetilde S}_0^\alpha M_0^{\alpha \beta} S_0^\beta - \nonumber \\
        &&\qquad \qquad - {1 \over 2} \, {\widetilde M}_0^{\alpha \beta} 
        \left( S_0^\alpha \nabla^2 S_0^\beta - S_0^\beta \nabla^2 S_0^\alpha 
                \right) \Biggr\} \ .
 \label{mctfun}
\end{eqnarray}
As usual, the harmonic part (\ref{harfun}) defines the propagators of the
field theory, while the perturbation expansion is performed in terms of the
non-linear vertices (\ref{relfun}) and (\ref{mctfun}).
Notice that the existence of the reversible forces (\ref{mctfun}) does not
show up in dynamic mean-field theory (van Hove theory), which in field-theory
language is based on the harmonic action (\ref{harfun}) only.

We can now construct the perturbation expansion for all possible correlation
functions of the dynamic and auxiliary fields, as well as for the associated 
vertex functions given by the one-particle irreducible Feynman diagrams.
A straightforward scaling analysis yields that the upper critical dimension of
this model is $d_c = 4$ for both the relaxational {\em and} the mode-coupling
vertices.
Therefore, for $d \leq 4$ the perturbation theory will be IR-singular, and 
non-trivial critical exponents will ensue, while for $d \geq 4$ the 
perturbation theory contains UV divergences.
In order to renormalize the field theory in the ultraviolet, it suffices to
render all the non-vanishing two-, three-, and four-point functions finite by
introducing multiplicative renormalization constants.
This is achieved by demanding the renormalized vertex functions, or appropriate
momentum and frequency derivatives thereof, to be finite when the fluctuation
integrals are taken at a conveniently chosen normalization point, well outside 
the IR regime.
We shall employ the dimensional regularization scheme in order to compute the
emerging momentum integrals, and choose the renormalized mass $\tau = 1$ as our
normalization point, or, sufficient to one-loop order, 
$\tau_0 = r_0 - r_{0c} = \mu^2$.
Notice that $\mu$ defines an intrinsic momentum scale of the renormalized 
theory.
The Gell-Mann--Low equation can subsequently be used to explore the dependence
of the {\em renormalized} correlation or vertex functions on $\mu$, and thereby
obtain information on their scaling behavior.

\subsubsection{Dynamics: Vertex function renormalization}

The UV-divergent two-, three-, and four-point vertex functions or their
derivatives that require multiplicative renormalization are
$\partial_\omega \Gamma_{0 \, {\widetilde M} M}({\bf q},\omega)$,
$\partial_{q^2} \Gamma_{0 \, {\widetilde M} M}({\bf q},\omega)$, and
$\partial_{q^2}$ $\Gamma_{0 \, {\widetilde M} {\widetilde M}}({\bf q},\omega)$;
$\Gamma_{0 \, {\widetilde S} S}({\bf q},\omega)$,
$\partial_\omega \Gamma_{0 \, {\widetilde S} S}({\bf q},\omega)$, 
$\partial_{q^2} \Gamma_{0 \, {\widetilde S} S}({\bf q},\omega)$, 

\noindent and 
$\partial_{q^2} \Gamma_{0 \, {\widetilde S} {\widetilde S}}({\bf q},\omega)$;
$\Gamma_{0 \, {\widetilde S} S M}(-{\bf q},-\omega;{{\bf q} \over 2}-{\bf p},
{\omega \over 2};{{\bf q} \over 2}+{\bf p},{\omega \over 2})$ and
$\partial_{({\bf q} \cdot {\bf p})} \Gamma_{0 \, {\widetilde M} S S}
(-{\bf q},-\omega; {{\bf q} \over 2}-{\bf p},{\omega \over 2};
{{\bf q} \over 2}+{\bf p},{\omega \over 2})$; and, at last,
$\Gamma_{0 \, {\widetilde S} S S S}(-{\bf q},-\omega;
{{\bf q} \over 3},{\omega \over 3}; {{\bf q} \over 3},{\omega \over 3};
{{\bf q} \over 3},{\omega \over 3})$.
On the other hand, we have four fluctuating fields 
(${\widetilde M}_0^{\alpha \beta}$, $M_0^{\alpha \beta}$, 
${\widetilde S}_0^\alpha$, $S_0^\alpha$) and the seven parameters 
${\widetilde D}_0$, $D_0$, ${\widetilde \lambda}_0$, $\lambda_0$, $\tau_0$, 
$g_0$, and $u_0$ available; this leaves us at liberty to choose one of the
renormalization constants in a convenient manner.

Starting with the two-point functions 
$\Gamma_{0 \, {\widetilde M} M}({\bf q},\omega)$ and 
$\Gamma_{0 \, {\widetilde S} S}({\bf q},\omega)$ for the conserved quantities 
and order parameter fluctuations, respectively, we immediately note that as a 
consequence of the momentum dependence of the mode-coupling vertices
\begin{equation}
        {\partial \over \partial (i \omega)} 
        \Gamma_{0 \, {\widetilde M} M}({\bf q} = {\bf 0},\omega) \equiv 1
 \label{amverf}
\end{equation}
to {\em all orders} in perturbation theory. 
Upon defining renormalized fields according to
\begin{eqnarray}
        &&{\widetilde M}^{\alpha \beta} = Z_{\widetilde M}^{1/2} 
                {\widetilde M}_0^{\alpha \beta} \; , \quad 
        M^{\alpha \beta} = Z_M^{1/2} M_0^{\alpha \beta} \ ,
 \label{conren} \\
        &&{\widetilde S}^\alpha = Z_{\widetilde S}^{1/2} 
                {\widetilde S}_0^\alpha \; , \quad 
        S^\alpha = Z_S^{1/2} S_0^\alpha \ ,
 \label{fldren}
\end{eqnarray}
and using $\Gamma_{{\widetilde M} M} = (Z_{\widetilde M} Z_M)^{-1/2} 
\Gamma_{0 \, {\widetilde M} M}$, we thus obtain the exact relation
\begin{equation}
        Z_{\widetilde M} Z_M \equiv 1 \ .
 \label{zmtrel}
\end{equation}
At this point we utilize our freedom of choice to set
\begin{equation}
        Z_{\widetilde M} \equiv Z_M \equiv 1 \ .
 \label{zmtcho}
\end{equation}

Similarly, for the order parameter fields we demand that $\partial 
\Gamma_{{\widetilde S} S}({\bf q} = {\bf 0},\omega) / \partial (i \omega)$ be
finite at the normalization point, which yields after evaluating the integrals
in dimensional renormalization \cite{fnote1},
\begin{eqnarray}
        &&(Z_{\widetilde S} Z_S)^{1/2} = 1 +
 \label{zstrel} \\
        &&\quad + {n-1 \over \epsilon} \left( 
          1 - {d_\parallel \over d} T_0^\parallel - {d_\perp \over d} T_0^\perp
          \right) {w_0 {\widetilde f}_0 A_d \mu^{-\epsilon} \over (1 + w_0)^2}
          \ . \nonumber
\end{eqnarray}
Here, $\epsilon = 4-d$, $A_d = \Gamma(3-d/2) / 2^{d-1} \pi^{d/2}$ is a 
geometric factor (non-singular near $d_c = 4$), $T_0^\parallel$ and $T_0^\perp$
denote the ratios defined in Eq.~(\ref{snoitm}), and the effective dynamic 
couplings are
\begin{eqnarray}
        &&w_0 = {\lambda_0 \over D_0} \ , 
 \label{defcpw} \\
        &&{\widetilde f}_0 = {{\widetilde \lambda}_0 \over \lambda_0} \, f_0 
                = {{\widetilde \lambda}_0 \over \lambda_0} \, 
                        {g_0^2 \over \lambda_0 D_0} \ .
 \label{defcpf}
\end{eqnarray}
Notice that in equilibrium, when $T_0^\parallel = T_0^\perp = 1$, 
Eq.~(\ref{zstrel}) yields $Z_{\widetilde S} Z_S^{1/2} = 1$ to this order.
In the isotropic noise case, with $T_0^\parallel = T_0^\perp$, or equivalently,
either $d_\parallel$ or $d_\perp = 0$, we recover the result cited in 
Ref.~\cite{uwezol}.
In the same way, all the subsequently found $Z$ factors reduce to the results 
for the isotropic non-equilibrium SSS model when 
$T_0^\parallel = T_0^\perp = T_0$, and these in turn to the well-established
equilibrium expressions for $T_0 = 1$.

As a next step, we compute the three-point vertex function 
$\Gamma_{0 \, {\widetilde S} S M}$ at zero external momenta and frequencies.
Upon defining the dimensionless renormalized mode-coupling constant
\begin{equation}
        g = Z_g^{1/2} \, g_0 \, A_d^{1/2} \, \mu^{-\epsilon/2} \ ,
 \label{mclren}
\end{equation}
this provides us with the product of $Z$ factors
\begin{eqnarray}
        &&(Z_{\widetilde S} Z_S Z_M Z_g)^{1/2} = 1 +
 \label{zsmrel} \\
        &&\quad + {n-1 \over \epsilon} \left( 
          1 - {d_\parallel \over d} T_0^\parallel - {d_\perp \over d} T_0^\perp
          \right) {w_0 {\widetilde f}_0 A_d \mu^{-\epsilon} \over (1 + w_0)^2}
          \ . \nonumber
\end{eqnarray}
Direct comparison with Eq.~(\ref{zstrel}) implies
\begin{equation}
        Z_g = Z_M^{-1} = Z_{\widetilde M} = 1 \ ,
 \label{zmtres}
\end{equation}
where the choice (\ref{zmtcho}) was employed.
As shown in the Appendix, as a consequence of the $O(n)$ invariance and the 
fact that the $M_0^{\alpha \beta}$ are the generators of the rotation symmetry 
group, one may derive a Ward identity leading to the {\em exact} relation
$Z_g Z_M \equiv 1$.
In equilibrium, this result is trivial, and a simple consequence of the fact
that the conserved fields $M_0^{\alpha \beta}$ are non-critical.
The absence of field renormalization therefore follows directly from the purely
quadratic appearance of the fields $M_0^{\alpha \beta}$ in the hamiltonian
(\ref{stasss}), which immediately implies that the static response function for
the generalized angular momenta is $X_0({\bf q},\omega=0) = 1$ exactly.
While this relation holds even in our variant of the SSS model with dynamic
anisotropy, see Eq.~(\ref{susidt}) below, one cannot directly infer the 
renormalization constant $Z_M$ therefrom, as there is no 
fluctuation-dissipation theorem to relate this response function with the 
corresponding correlation function in the general non-equilibrium situation. 
Therefore $Z_g$ needs to be computed explicitly from the three-point vertex
function, or inferred from the above-mentioned Ward identity.

The other vertex function renormalizing the mode-coupling constant $g_0$ is 
$\Gamma_{0 \, {\widetilde M} S S}(-{\bf q},-\omega;{{\bf q} \over 2}-{\bf p},
{\omega \over 2};{{\bf q} \over 2}+{\bf p},{\omega \over 2})$ 
$= 2 g_0 ({\bf q} \cdot {\bf p}) + O(g_0^3)$.
Thus, in order to obtain $Z_S$, we need to take a derivative with respect to
$({\bf q} \cdot {\bf p})$; but owing to the dynamical anisotropy, the result
depends on whether the components $q_i$ and $p_i$ lie in the longitudinal or
transverse sector in momentum space, respectively.
This means that we have to introduce {\em different} field renormalizations
$Z_S^\parallel$ and $Z_S^\perp$ for the longitudinal and transverse field 
fluctuations, for which one then finds to one-loop order
\begin{eqnarray}
        &&Z_S^{\parallel / \perp} = 1 +
 \label{zspatr} \\
        &&\quad + {n-1 \over 2 \epsilon} \left( 
          1 - {d_\parallel \over d} T_0^\parallel - {d_\perp \over d} T_0^\perp
          \right) {{\widetilde f}_0 A_d \mu^{-\epsilon} \over (1 + w_0)^2} \mp
        \nonumber \\
        &&\quad \mp {n-1 \over 6 \epsilon} {d_{\perp / \parallel} \over d} 
        \left( T_0^\parallel - T_0^\perp \right) {1 + 2 w_0 \over (1 + w_0)^2}
        \, {\widetilde f}_0 A_d \mu^{-\epsilon} \ . \nonumber
\end{eqnarray}
Obviously, for $T_0^\parallel = T_0^\perp$, this novel distinction between
$Z_S^\parallel$ and $Z_S^\perp$ disappears, and the isotropic result of 
Ref.~\cite{uwezol} is recovered.
Combining Eq.~(\ref{zspatr}) with Eq.~(\ref{zstrel}) yields
\begin{eqnarray}
        &&Z_{\widetilde S}^{\parallel / \perp} = 1 -
 \label{zstptr} \\
        &&\; - {n-1 \over 2 \epsilon} \left( 
          1 - {d_\parallel \over d} T_0^\parallel - {d_\perp \over d} T_0^\perp
          \right) {1 - 4 w_0 \over (1 + w_0)^2} \, 
        {\widetilde f}_0 A_d \mu^{-\epsilon} \pm \nonumber \\
        &&\quad \pm {n-1 \over 6 \epsilon} {d_{\perp / \parallel} \over d} 
        \left( T_0^\parallel - T_0^\perp \right) {1 + 2 w_0 \over (1 + w_0)^2}
        \, {\widetilde f}_0 A_d \mu^{-\epsilon} \ . \nonumber
\end{eqnarray}

Next we define renormalized transport coefficients and noise strengths 
according to \cite{fnote2}
\begin{eqnarray}
        &&{\widetilde D}^{\parallel/\perp} = Z_{\widetilde D}^{\parallel/\perp}
        \, {\widetilde D}_0^{\parallel/\perp} \; , \quad 
        D^{\parallel/\perp} = Z_D^{\parallel/\perp} \, D_0 \ , 
 \label{noiren} \\
        &&{\widetilde \lambda}^{\parallel/\perp} = 
        Z_{\widetilde \lambda}^{\parallel/\perp} \, {\widetilde \lambda}_0 \; ,
        \quad \lambda^{\parallel/\perp} = 
        Z_\lambda^{\parallel/\perp} \, \lambda_0 \ , 
 \label{onsren}
\end{eqnarray}
where we allow for {\em different} renormalizations in the parallel and 
transverse sectors of the originally isotropic parameters $D_0$,
${\widetilde \lambda}_0$, and $\lambda_0$.
The renormalized noise coefficients can be obtained by demanding that the
vertex functions $\partial_{q_{\parallel/\perp}^2} 
\Gamma_{{\widetilde M}{\widetilde M}}({\bf q}_{\parallel/\perp},\omega=0) 
\vert_{{\bf q}_{\parallel/\perp}={\bf 0}}$ and 
$\Gamma_{{\widetilde S}{\widetilde S}}({\bf q}={\bf 0},\omega=0)$ be UV-finite.
This yields, with $Z_{\widetilde M} = 1$,
\begin{equation}
        Z_{\widetilde D}^{\parallel/\perp} = 1 + {1 \over 2 \epsilon} \,
        {{\widetilde f}_0 A_d \mu^{-\epsilon} \over T_0^{\parallel/\perp}} \ ,
 \label{zdtptr}
\end{equation}
and
\begin{equation}
        Z_{\widetilde S} \, Z_{\widetilde \lambda} = 
        1 + {n-1 \over \epsilon} \left( {d_\parallel \over d} T_0^\parallel 
                + {d_\perp \over d} T_0^\perp \right) 
                {{\widetilde f}_0 A_d \mu^{-\epsilon} \over 1 + w_0} \ .
 \label{zltptr}
\end{equation}
While this product is still isotropic, the anisotropy in the field 
renormalization (\ref{zstptr}) induces different renormalized order parameter 
noise strengths in the longitudinal and transverse sectors.

In the same manner, from $\partial_{q^2}
\Gamma_{{\widetilde M}M}({\bf q},\omega=0) \vert_{{\bf q}={\bf 0}}$ and 
$\partial_{q_{\parallel/\perp}^2} 
\Gamma_{{\widetilde S}S}({\bf q}_{\parallel/\perp},\omega=0) 
\vert_{{\bf q}_{\parallel/\perp}={\bf 0}}$ we obtain
\begin{equation}
        Z_D = 1 + {1 \over 2 \epsilon} \, {\widetilde f}_0 A_d \mu^{-\epsilon}
        \ ,
 \label{zdpatr}
\end{equation}
and
\begin{eqnarray}
        &&(Z_{\widetilde S} Z_S)^{1/2} \, Z_\lambda^{\parallel/\perp} =
        1 + {n-1 \over \epsilon} \, 
        {{\widetilde f}_0 A_d \mu^{-\epsilon} \over 1 + w_0} +
 \label{zlpatr} \\
        &&\qquad + {n-1 \over \epsilon} \left( 
        1 - {d_\parallel \over d} T_0^\parallel - {d_\perp \over d} T_0^\perp 
        \right) {w_0 {\widetilde f}_0 A_d \mu^{-\epsilon} \over (1 + w_0)^3} 
        \mp \nonumber \\
        &&\qquad \mp {n-1 \over 3 \epsilon} {d_{\perp / \parallel} \over d} 
        \left( T_0^\parallel - T_0^\perp \right) {w_0^2 {\widetilde f}_0 A_d 
                \mu^{-\epsilon} \over (1 + w_0)^3} \ . \nonumber
\end{eqnarray}
Thus, to one-loop order at least, we need not distinguish between the
renormalized diffusion constants $D^\parallel$ and $D^\perp$.
This concludes our multiplicative vertex function renormalization for the
dynamical parameters of the non-equili\-brium SSS model with dynamical 
anisotropy.
Notice that the anisotropic parts of the renormalization constants are always
proportional to $T_0^\parallel - T_0^\perp$.
For $T_0^\parallel = T_0^\perp = 1$, we recover the equilibrium results where
$Z_{\widetilde D} = Z_D$ and $Z_{\widetilde \lambda} = Z_\lambda$, reflecting 
the Einstein relation.

\subsubsection{Statics: Response function renormalization}

In order to define the ``static'' limit of the intrinsically dynamic model 
under consideration here, we compute the response functions for the generalized
angular momenta and the order parameter components, and then take the limit
$\omega \to 0$ there.
By adding external fields to the hamiltonian (\ref{stasss}), one may show that
the dynamic susceptibilities for the conserved quantities $M_0^{\alpha \beta}$
and the order parameter fluctuations $S_0^{\alpha}$ are given by
\begin{eqnarray}
        X_0({\bf q},\omega) &=& 
        \Gamma_{0\, {\widetilde M}M}(-{\bf q},-\omega)^{-1} \times \nonumber \\
        &&\times \left[ D_0 q^2 + 2 g_0 \,
        \Gamma_{0\, {\widetilde M}[{\widetilde S}S]}(-{\bf q},-\omega) 
        \right] ,
 \label{amvsus} \\
        \chi_0({\bf q},\omega) &=& 
        \Gamma_{0\, {\widetilde S}S}(-{\bf q},-\omega)^{-1} \times \nonumber \\
        &&\times \left[ \lambda_0 - g_0 \,
        \Gamma_{0\, {\widetilde S}[{\widetilde M}S]}(-{\bf q},-\omega) 
        \right] \ ,
 \label{opvsus}
\end{eqnarray}
respectively \cite{dynfun,uwezol}.
Notice that composite-operator vertex functions enter these expressions, which
in general implies that new renormalization constants are required to remove 
the UV singularities of the response functions (equivalently, one may utilize
the $Z$ factors obtained from the multiplicative renormalization of the vertex
functions plus appropriate additive renormalizations \cite{dynfun,uwezol}).

Yet one may show to {\em all orders} in perturbation theory that
\begin{equation}
        \Gamma_{0\, {\widetilde M}M}({\bf q},\omega) = i \omega + D_0 q^2 + 2 
        g_0 \, \Gamma_{0\, {\widetilde M}[{\widetilde S}S]}({\bf q},\omega) \ ,
 \label{susidt}
\end{equation}
and consequently
\begin{equation}
        X_0({\bf q},\omega=0) \equiv 1 \ , 
 \label{stamsu}
\end{equation}
which means that there is no additional renormalization required here.
On the other hand, the static limit of the order parameter susceptibility is in
fact singular, which leads us to define the corresponding renormalized response
function via
\begin{equation}
        \chi({\bf q},\omega) = Z \, \chi_0({\bf q},\omega) \ .
 \label{rensus} 
\end{equation}
The new renormalization constant $Z$ is determined by demanding that 
$\partial_{q_{\parallel/\perp}^2} \chi({\bf q}_{\parallel/\perp},\omega=0)^{-1}
\vert_{{\bf q}_{\parallel/\perp}={\bf 0}}$ be UV-finite; once again, the result
is different in the longitudinal and transverse momentum space sectors:
\begin{eqnarray}
        Z^{\parallel / \perp} &=& 1 + {n-1 \over \epsilon} \left( 
        1 - {d_\parallel \over d} T_0^\parallel - {d_\perp \over d} T_0^\perp 
        \right) {w_0 {\widetilde f}_0 A_d \mu^{-\epsilon} \over (1 + w_0)^3} 
        \mp \nonumber \\
        &&\mp {n-1 \over 3 \epsilon} {d_{\perp / \parallel} \over d} 
        \left( T_0^\parallel - T_0^\perp \right) {w_0^2 {\widetilde f}_0 A_d 
                \mu^{-\epsilon} \over (1 + w_0)^3} \ .
 \label{zetptr}
\end{eqnarray}
To one-loop order, $Z^{\parallel / \perp}$ as well as $Z_S^{\parallel / \perp}$
and $Z_{\widetilde S}^{\parallel / \perp}$, do not contain the static 
non-linear coupling $u_0$.

This is different for the remaining two $Z$ factors needed for the renormalized
dimensionless distance from the critical point $\tau$ and the static 
non-linearity $u$ \cite{fnote2},
\begin{eqnarray}
        &&\tau^{\parallel/\perp} = Z_\tau^{\parallel/\perp} \, \tau_0 \, 
                \mu^{-2} \; , \quad \tau_0 = r_0 - r_{0c} \ ,
 \label{tauren} \\
        &&u^{\parallel/\perp} = Z_u^{\parallel/\perp} \, u_0 \, A_d \, 
                \mu^{- \epsilon} \ . 
 \label{sturen}
\end{eqnarray} 
The fluctuation-induced $T_c$ shift is determined from the criticality
condition $\chi_0({\bf q} = {\bf 0},\omega = 0)^{-1} = 0$ at $r_0 = r_{0c}$
($\tau_0 = 0$) with the result
\begin{eqnarray}
        &&r_{0c} = - {n+2 \over 6} \, {\widetilde u}_0 
                \int_k \! {1 \over r_{0c} + k^2} + (n-1) \times
 \label{tcsint} \\
        &&\quad \times \left( 1 - {d_\parallel \over d} T_0^\parallel 
        - {d_\perp \over d} T_0^\perp \right){\widetilde f}_0 
                \int_k \! {1 \over w_0 (r_{0c} + k^2) + k^2} \ . \nonumber
\end{eqnarray}
In principle, these momentum integrals should be evaluated with a {\em finite} 
upper cutoff, which underlines the non-universality of the $T_c$ shift, i.e.,
its dependence on short-distance properties.
However, if we choose to evaluate the momentum integrals by means of 
dimensional regularization, we are led to  
\begin{eqnarray}
        &&|r_{0c}| = \Biggl( {2 A_d \over (d-2) (4-d)} 
        \Biggr[ {n+2 \over 6} \, {\widetilde u}_0 - {n-1 \over 1 + w_0} \times
\label{tcshif} \\
        &&\quad \times \left( {w_0 \over 1 + w_0} \right)^{{d\over 2}-1}
        \left( 1 - {d_\parallel \over d} T_0^\parallel - {d_\perp \over d} 
        T_0^\perp \right){\widetilde f}_0 \Biggr] \Biggr)^{2 \over 4-d} \ ,
        \nonumber
\end{eqnarray} 
where we have defined
\begin{equation}
        {\widetilde u}_0 = {{\widetilde \lambda}_0 \over \lambda_0} \, u_0 \ .
 \label{effcpu}
\end{equation}
Rendering $\chi({\bf q}={\bf 0},\omega=0)^{-1}$ UV-finite, after substituting 
$r_0 = \tau_0 + r_{0c}$, then yields the isotropic product
\begin{eqnarray}
        &&Z \, Z_\tau = 1 - {n+2 \over 6 \epsilon} \, 
                {\widetilde u}_0 A_d \mu^{- \epsilon} +
 \label{ztaupt} \\
        &&\qquad + {n-1 \over \epsilon} \left( 1 - {d_\parallel \over d} 
                T_0^\parallel - {d_\perp \over d} T_0^\perp \right) {w_0 
        {\widetilde f}_0 A_d \mu^{-\epsilon} \over (1 + w_0)^2} \ . \nonumber
\end{eqnarray}
The anisotropy in the $Z^{\parallel/\perp}$ from Eq.~(\ref{zetptr}) then 
induces the different $Z$ factors for the longitudinal and transverse momentum 
space sectors anticipated in Eq.~(\ref{tauren}).
We note that alternatively we could have used 
$\Gamma_{{\widetilde S}S}({\bf 0},0)$, providing the (isotropic) combination 
$(Z_{\widetilde S} Z_S)^{1/2} Z_\lambda Z_\tau$.
(Indeed, the anisotropic contributions to $Z^{\parallel/\perp}$ 
[Eq.~(\ref{zetptr})] and to $Z_\lambda^{\parallel/\perp}$ [Eq.~(\ref{zlpatr})]
are identical.)

Finally, we need the renormalization of the static coupling $u_0$, which we may
obtain from the four-point function $\Gamma_{{\widetilde S} S S S}$ at 
vanishing external wavevectors and frequencies.
To one-loop order, there appear ten Feynman diagrams, the contributions of nine
of which cancel in equilibrium (and {\em only} there!).
A somewhat tedious calculation eventually gives
\begin{eqnarray}
        &&(Z_{\widetilde S} Z_S)^{1/2} \, Z_S \,  Z_\lambda \, Z_u = 
 \label{zupatr} \\
        &&\quad 1 - {n+8 \over 6 \epsilon} \, {\widetilde u}_0 A_d 
                \mu^{- \epsilon} + {n-1 \over \epsilon} \, {{\widetilde f}_0 
                A_d \mu^{- \epsilon} \over 1 + w_0} - \nonumber \\
        &&\qquad - {n-1 \over \epsilon} \left( 1 - {d_\parallel \over d} 
        T_0^\parallel - {d_\perp \over d} T_0^\perp \right) {{\widetilde f}_0 
                A_d \mu^{-\epsilon} \over (1 + w_0)^2} - \nonumber \\
        &&\qquad + {n-1 \over \epsilon} \left( 1 - {d_\parallel \over d} 
        T_0^\parallel - {d_\perp \over d} T_0^\perp \right) 
        {6 {\widetilde f}_0^2 A_d \mu^{-\epsilon} \over (1 + w_0) 
                {\widetilde u}_0} \ . \nonumber
\end{eqnarray}
The anisotropies in $Z_S^{\parallel/\perp} Z_\lambda^{\parallel/\perp}$ again
cause the differences in $Z_u^\parallel$ and $Z_u^\perp$.
No further UV renormalizations are required, and we may now turn to the
analysis of the ensuing RG flow equations.

\subsection{Discussion of the RG flow equations}
 \label{sssflo}

\subsubsection{RG equations for the vertex and response functions}

By means of the above renormalization constants, we can now write down the RG
(Gell-Mann--Low) equations for the vertex functions and the dynamic
susceptibilities. 
The latter connect the asymptotic theory, where the IR singularities become 
manifest, with a region in parameter space where the loop integrals are finite 
and ordinary ``naive'' perturbation expansion is applicable, and follow from 
the simple observation that the ``bare'' vertex functions do not depend on the 
renormalization scale $\mu$,
\begin{equation}
        \mu {d \over d\mu} \bigg\vert_0 \Gamma_{0 \, {\widetilde M}^k 
        {\widetilde S}^r M^l S^s}(\{ {\bf q},\omega \} ; \{ a_0 \}) = 0 \ ,
 \label{rngreq}
\end{equation}
with $\{ a_0 \} = g_0$,${\widetilde D}_0^\parallel$,${\widetilde D}_0^\perp$,
$D_0$,${\widetilde \lambda}_0$,$\lambda_0$,$\tau_0$,$u_0$.
For the non-equi\-librium SSS model with dynamical anisotropy, we have to treat
the longitudinal and transverse sectors in momentum space {\em separately}, 
i.e., we need to understand the flow of the renormalized set of parameters 
$\{ a^\parallel \} = g$,${\widetilde D}^\parallel$,$D^\parallel$,
${\widetilde \lambda}^\parallel$,$\lambda^\parallel$,$\tau^\parallel$,
$u^\parallel$, and $\{ a^\perp \} = g$,${\widetilde D}^\perp$,$D^\perp$, 
${\widetilde \lambda}^\perp$,$\lambda^\perp$,$\tau^\perp$,$u^\perp$, 
respectively.
Replacing the bare parameters and fields in Eq.~(\ref{rngreq}) with the
renormalized ones, we thus find the following partial differential equations
for the renormalized vertex functions in the longitudinal and transverse 
sectors,
\begin{eqnarray}
        &&\left[ \mu {\partial \over \partial \mu} 
        + \sum_{\{ a^{\parallel/\perp} \}} \zeta_a^{\parallel/\perp}
        a^{\parallel/\perp} {\partial \over \partial a^{\parallel/\perp}} 
        + {r \over 2} \, \zeta_{\widetilde S}^{\parallel/\perp} 
        + {s \over 2} \, \zeta_S^{\parallel/\perp} \right] \times \nonumber \\ 
        &&\quad \times \Gamma_{{\widetilde M}^k {\widetilde S}^r M^l S^s} 
        \left( \mu, \{ {\bf q}_{\parallel/\perp},\omega \};
                \{ a^{\parallel/\perp} \} \right) = 0 \ .
 \label{calsym}
\end{eqnarray}
Here, we have introduced Wilson's flow functions
\begin{eqnarray}
        \zeta_{\widetilde S}^{\parallel/\perp} &=& 
                \mu {\partial \over \partial \mu} \bigg\vert_0 
                \ln Z_{\widetilde S}^{\parallel/\perp} \ ,
 \label{zetflt} \\
        \zeta_S^{\parallel/\perp} &=& \mu {\partial \over \partial \mu} 
                \bigg\vert_0 \ln Z_S^{\parallel/\perp} \ ,
 \label{zetfld}
\end{eqnarray}
and
\begin{equation}
        \zeta_a^{\parallel/\perp} = \mu {\partial \over \partial \mu}
                \bigg\vert_0 \ln {a^{\parallel/\perp} \over a_0}
 \label{zetpar}
\end{equation}
(the index ``0'' indicates that the renormalized fields and parameters are to 
be expressed in terms of their bare counterparts prior to performing the
derivatives with respect to the momentum scale $\mu$).
Note that $\zeta_{\widetilde M} = \zeta_M \equiv 0$ and 
$\zeta_g \equiv - \epsilon / 2$ as a consequence of Eqs.~(\ref{zmtres}) and
(\ref{mclren}).

The Gell-Mann--Low equation (\ref{calsym}) is readily solved with the method 
of characteristics $\mu \to \mu \ell$; this defines running couplings
as the solutions to the first-order differential RG flow equations
\begin{equation}
        \ell \, {d a^{\parallel/\perp}(\ell) \over d\ell} = 
        \zeta_a^{\parallel/\perp}(\ell) \, a^{\parallel/\perp}(\ell) \, , \
        a^{\parallel/\perp}(1) = a^{\parallel/\perp} \ .      
 \label{rgflow}
\end{equation}
The solutions of the partial differential equations (\ref{calsym}) then read
\begin{eqnarray}
        &&\Gamma_{{\widetilde M}^k {\widetilde S}^r M^l S^l} \left( \mu, \{ 
        {\bf q}_{\parallel/\perp},\omega \};\{ a^{\parallel/\perp} \} \right) =
 \label{solcsy} \\
        &&\quad = \exp \left\{ {1 \over 2} \int_1^\ell 
        \Bigl[ r \, \zeta_{\widetilde S}^{\parallel/\perp}(\ell') 
        + s \, \zeta_S^{\parallel/\perp}(\ell') \Bigr] {d \ell' \over \ell'} 
                \right\} \times \nonumber \\ 
        &&\quad \times \Gamma_{{\widetilde M}^k {\widetilde S}^r M^l S^s} 
        \left( \mu \ell, \{ {\bf q}_{\parallel/\perp},\omega \};
                \{ a^{\parallel/\perp}(\ell) \} \right) \ . \nonumber
\end{eqnarray}
In the same manner, one can solve the RG equations for the dynamic 
susceptibilities, with the results
\begin{eqnarray}
        &&X \left( \mu, \{ {\bf q}_{\parallel/\perp},\omega \};
                \{ a^{\parallel/\perp} \} \right) = 
 \label{solcsx} \\
        &&\quad = X \left( \mu \ell, \{ {\bf q}_{\parallel/\perp},\omega \};
                \{ a^{\parallel/\perp}(\ell) \} \right) \ , \nonumber
\end{eqnarray}
and
\begin{eqnarray}
        &&\chi \left( \mu, \{ {\bf q}_{\parallel/\perp},\omega \};
                \{ a^{\parallel/\perp} \} \right) = 
 \label{solcsc} \\
        &&\qquad = \exp \left\{ - \int_1^\ell \zeta^{\parallel/\perp}(\ell') 
        {d \ell' \over \ell'} \right\} \times \nonumber \\
        &&\qquad \quad \times \chi \left( \mu \ell, 
        \{ {\bf q}_{\parallel/\perp},\omega \};\{ a^{\parallel/\perp}(\ell) \} 
                \right) \ , \nonumber
\end{eqnarray}
where, in analogy with Eq.~(\ref{zetfld}),
\begin{equation}
        \zeta^{\parallel/\perp} = \mu {\partial \over \partial \mu} 
                \bigg\vert_0 \ln Z^{\parallel/\perp} \ .
 \label{zetflc}
\end{equation}

Upon introducing renormalized anisotropic counterparts for the effective 
dynamic couplings (\ref{defcpw}), (\ref{defcpf}), and (\ref{snoitm}),
\begin{eqnarray}
        &&w^{\parallel/\perp} = 
                {\lambda^{\parallel/\perp} \over D^{\parallel/\perp}} \ , 
 \label{defcrw} \\
        &&{\widetilde f}^{\parallel/\perp} = 
        {{\widetilde \lambda}^{\parallel/\perp}\over \lambda^{\parallel/\perp}}
        \, {g^2 \over \lambda^{\parallel/\perp} D^{\parallel/\perp}} \ ,
 \label{defcrf} \\
        &&T^{\parallel/\perp} = 
        {{\widetilde D}^{\parallel/\perp} \over D^{\parallel/\perp}} \, 
        {\lambda^{\parallel/\perp}\over {\widetilde \lambda}^{\parallel/\perp}}
                \ ,
 \label{defcrt}
\end{eqnarray}
the zeta functions to one-loop order read explicitly
\begin{eqnarray}
        &&\zeta_{\widetilde S}^{\parallel/\perp} = {n-1 \over 2} \left( 1 
        - {d_\parallel \over d} T^\parallel - {d_\perp \over d} T^\perp \right)
        {(1 - 4 w^{\parallel/\perp}) \, {\widetilde f}^{\parallel/\perp} \over 
        (1 + w^{\parallel/\perp})^2} \nonumber \\
        &&\quad \mp {n-1 \over 6} \, {d_{\perp/\parallel} \over d} 
        \left( T^\parallel - T^\perp \right) {(1 + 2 w^{\parallel/\perp}) \,
        {\widetilde f}^{\parallel/\perp} \over (1 + w^{\parallel/\perp})^2} \ ,
 \label{zestpt} \\
        &&\zeta_S^{\parallel/\perp} = - {n-1 \over 2} \left( 1 
        - {d_\parallel \over d} T^\parallel - {d_\perp \over d} T^\perp \right)
        {{\widetilde f}^{\parallel/\perp} \over (1 + w^{\parallel/\perp})^2} 
                \pm \nonumber \\
        &&\quad \pm {n-1 \over 6} \, {d_{\perp/\parallel} \over d} 
        \left( T^\parallel - T^\perp \right) {(1 + 2 w^{\parallel/\perp}) \,
        {\widetilde f}^{\parallel/\perp} \over (1 + w^{\parallel/\perp})^2} \ ,
 \label{zesptr} \\
        &&\zeta^{\parallel/\perp} = - (n-1) \left( 1 - {d_\parallel \over d} 
        T^\parallel - {d_\perp \over d} T^\perp \right) {w^{\parallel/\perp} \,
        {\widetilde f}^{\parallel/\perp} \over (1 + w^{\parallel/\perp})^3} \mp
                \nonumber \\
        &&\quad \mp {n-1 \over 3} \, {d_{\perp/\parallel} \over d} 
        \left( T^\parallel - T^\perp \right) {(w^{\parallel/\perp})^2 \,
        {\widetilde f}^{\parallel/\perp} \over (1 + w^{\parallel/\perp})^3} \ ,
 \label{zepatr} \\
        &&\zeta_g \equiv - {\epsilon \over 2} \ ,
 \label{zetage} \\
        &&\zeta_{\widetilde D}^{\parallel/\perp} = - {1 \over 2} \, 
                {{\widetilde f}^{\parallel/\perp} \over T^{\parallel/\perp}}\ ,
 \label{zedtpt} \\
        &&\zeta_{\widetilde \lambda}^{\parallel/\perp} = - (n-1) 
        {{\widetilde f}^{\parallel/\perp} \over 1 + w^{\parallel/\perp}} +
                \nonumber \\
        &&\quad + {n-1 \over 2} \left( 1 - {d_\parallel \over d} T^\parallel 
        - {d_\perp \over d} T^\perp \right) {(1 + 6 w^{\parallel/\perp}) \,
        {\widetilde f}^{\parallel/\perp} \over (1 + w^{\parallel/\perp})^2} \mp
                \nonumber \\
        &&\quad \mp {n-1 \over 6} \, {d_{\perp/\parallel} \over d} 
        \left( T^\parallel - T^\perp \right) {(1 + 2 w^{\parallel/\perp}) \,
        {\widetilde f}^{\parallel/\perp} \over (1 + w^{\parallel/\perp})^2} \ ,
 \label{zelspt} \\
        &&\zeta_D^{\parallel/\perp} = 
        - {1 \over 2} \, {\widetilde f}^{\parallel/\perp} \ ,
 \label{zetdpt} \\
        &&\zeta_\lambda^{\parallel/\perp} = - (n-1) 
        {{\widetilde f}^{\parallel/\perp} \over 1 + w^{\parallel/\perp}} +
                \nonumber \\
        &&\quad + (n-1) \left( 1 - {d_\parallel \over d} T^\parallel 
        - {d_\perp \over d} T^\perp \right) {(w^{\parallel/\perp})^2 \, 
        {\widetilde f}^{\parallel/\perp} \over (1 + w^{\parallel/\perp})^3} \mp
                \nonumber \\
        &&\quad \mp {n-1 \over 3} \, {d_{\perp/\parallel} \over d} 
        \left( T^\parallel - T^\perp \right) {(w^{\parallel/\perp})^2 \,
        {\widetilde f}^{\parallel/\perp} \over (1 + w^{\parallel/\perp})^3} \ ,
 \label{zeltpt} \\                
        &&\zeta_\tau^{\parallel/\perp} = - 2 
                + {n+2 \over 6} \, {\widetilde u}^{\parallel/\perp} -
                \nonumber \\
        &&\quad - (n-1) \left( 1 - {d_\parallel \over d} T^\parallel 
        - {d_\perp \over d} T^\perp \right) {{w^{\parallel/\perp}}^2 \,
        {\widetilde f}^{\parallel/\perp} \over (1 + w^{\parallel/\perp})^3} \mp
                \nonumber \\
        &&\quad \mp {n-1 \over 3} \, {d_{\perp/\parallel} \over d} 
        \left( T^\parallel - T^\perp \right) {{w^{\parallel/\perp}}^2 \,
        {\widetilde f}^{\parallel/\perp} \over (1 + w^{\parallel/\perp})^3} \ ,
 \label{zetrpt} \\      
        &&\zeta_u^{\parallel/\perp} = - \epsilon 
                + {n+2 \over 8} \, {\widetilde u}^{\parallel/\perp} + 
 \label{zetupt} \\
        &&\quad + {n-1 \over 2} \left( 1 - {d_\parallel \over d} T^\parallel 
        - {d_\perp \over d} T^\perp \right) {(3 + 5 w^{\parallel/\perp}) \, 
        {\widetilde f}^{\parallel/\perp} \over (1 + w^{\parallel/\perp})^3} -
                \nonumber \\
        &&\quad - (n-1) \left( 1 - {d_\parallel \over d} T^\parallel 
        - {d_\perp \over d} T^\perp \right) 
        {6 ({\widetilde f}^{\parallel/\perp})^2 \over (1 + w^{\parallel/\perp})
                \, {\widetilde u}^{\parallel/\perp}} \pm \nonumber \\
        &&\pm {n-1 \over 6} \, {d_{\perp/\parallel} \over d} 
        \left( T^\parallel - T^\perp \right) {(1 + 3 w^{\parallel/\perp} 
        + 4 {w^{\parallel/\perp}}^2) {\widetilde f}^{\parallel/\perp} \over 
                (1 + w^{\parallel/\perp})^3} \ . \nonumber 
\end{eqnarray}
These results enable us now to study the scaling behavior of the 
non-equilibrium SSS model with dynamical noise in the vicinity of the different
RG fixed points, which are given by the zeros of the appropriate RG beta 
functions
\begin{equation}
        \beta_v = \mu {\partial \over \partial \mu} \bigg\vert_0 v \ .
 \label{defbet}
\end{equation}
According to
\begin{equation}
        \ell \, {d v(\ell) \over d \ell} = \beta_v(\{ v(\ell) \}) \ ,
 \label{cpflow}
\end{equation}
these govern the flow of the effective couplings $T^{\parallel/\perp}$, 
$w^{\parallel/\perp}$, ${\widetilde f}^{\parallel/\perp}$, and
${\widetilde u}^{\parallel/\perp}$, etc. under scale transformations 
$\mu \to \mu \ell$, and the fixed points $\{ v^* \}$ where all 
$\beta_v(\{ v^* \}) = 0$ thus describe scale-invariant regimes.

\subsubsection{RG fixed points and critical exponents}

\begin{figure}
  \centerline{\epsfxsize 0.75\columnwidth \epsfbox{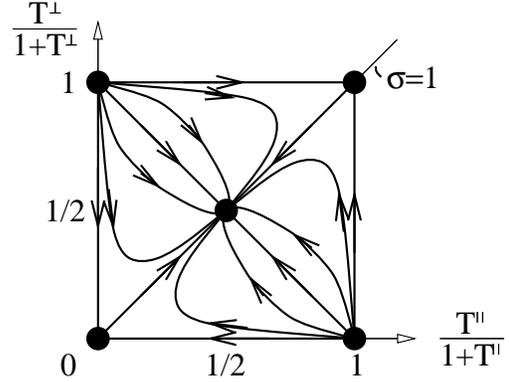}}
\caption{Equilibrium (center) and non-equilibrium (corners) fixed points of the
  non-equilibrium SSS model with dynamical anisotropy in the 
  [$T^\parallel/(1+T^\parallel),T^\perp/(1+T^\perp)$]--plane.}
 \label{sssfig}
\end{figure}

We begin with considering the RG flow of the anisotropy parameter
\begin{equation}
        \sigma = T^\perp / T^\parallel \ ,
 \label{defsig}
\end{equation}
denoting the ratio of the effective conserved noise temperatures (\ref{defcrt})
in the transverse and longitudinal sectors ($0 \leq \sigma \leq 1$).
Obviously, $\sigma = 1$ describes the isotropic fixed point. 
In order to assess its stability against the anisotropic non-equilibrium
perturbation, we consider the RG beta function
\begin{eqnarray}
        &&\beta_\sigma = \mu {\partial \over \partial \mu}\bigg\vert_0 \sigma =
  \label{betasg} \\
        &&\quad = \sigma \left( 
                \zeta_{\widetilde D}^\perp - \zeta_{\widetilde D}^\parallel 
                - \zeta_D^\perp + \zeta_D^\parallel + \zeta_\lambda^\perp 
                - \zeta_\lambda^\parallel - \zeta_{\widetilde \lambda}^\perp 
                + \zeta_{\widetilde \lambda}^\parallel \right) \ . \nonumber
\end{eqnarray}
As to first order $w^\parallel = w^\perp = w$ and 
${\widetilde f}^\perp = {\widetilde f}^\parallel = {\widetilde f}$ (in the
vicinity of the isotropic fixed point, this holds even beyond the one-loop
approximation), we may write
\begin{equation}
        \beta_\sigma = - \sigma (1 - \sigma) {\widetilde f} T^\parallel 
        \left[ {1 \over 2 T^\parallel T^\perp} 
                + {n-1 \over 6} \, {1 + 3 w \over (1 + w)^3} \right] \ .
 \label{betsig}
\end{equation}
The expression in square brackets is positive, and thus, as to be expected on 
physical grounds, there are only two fixed points, namely $\sigma^* = 1$ and 
$\sigma^* = 0$, realized for $T^\perp = 0$ and $T^\parallel = \infty$. 
(Of course, if we allow for $\sigma > 1$ as well, there is also the fixed point
$\sigma^* = \infty$, realized for $T^\parallel = 0$ and $T^\perp = \infty$; yet
clearly the regimes $\sigma < 1$ and $\sigma > 1$ map onto each other through
simply relabeling the indices $\parallel \leftrightarrow \perp$.)
Furthermore, in the IR regime ($\ell \to 0$), if $0 < \sigma < 1$ initially,
$\beta_\sigma < 0$, and thus $\sigma(\ell)$ grows until it reaches the 
isotropic fixed point $\sigma^* = 1$ (and conversely, if $1 < \sigma < \infty$,
then $\beta_\sigma > 0$ and $\sigma(\ell)$ decreases towards $\sigma^* = 1$).
Thus, the {\em isotropic} fixed point is {\em stable} against the spatially
anisotropic perturbations in the noise correlator of the conserved generalized
angular momenta.
Figure~\ref{sssfig} depicts these various fixed points in the non-equilibrium
SSS model with dynamical anisotropy, and the parameter flow in the
[$T^\parallel/(1+T^\parallel),T^\perp/(1+T^\perp)$]--plane.
The center of this diagram represents the equilibrium SSS dynamic fixed point
($\sigma = 1$, $T = 1$), and is attractive for the RG flows originating from 
any point inside the depicted square.
Precisely on the edges in parameter space, we find the two isotropic 
non-equilibrium fixed points with $\sigma = 1$ and $T = 0$ in the lower left, 
and $T = \infty$ in the upper right corners, respectively, and the two
anisotropic fixed points with $\sigma = 0$ and $\sigma = \infty$ in the lower
right and upper left corners.
Notice that the RG flows on this {\em critical} surface in parameter space
which start in the vicinity of these anisotropic non-equilibrium fixed points 
tend towards the isotropic non-equilibrium fixed points, but eventually end up 
at the isotropic equilibrium fixed point provided that initially 
$0 < T^\parallel < \infty$ and $0 < T^\perp < \infty$.  

Before we investigate the properties of the anisotropic fixed point 
$\sigma^* = 0$ ($T^\perp = 0$, $T^\parallel = \infty$), let us briefly 
summarize the behavior of the isotropic model with $\sigma^* = 1$.
(For more details, and for a graph depicting the various equilibrium and 
non-equilibrium fixed points in the isotropic parameter subspace, we refer to
Ref.~\cite{uwezol}.)
Setting $T^\parallel = T^\perp = T$ in the flow functions 
(\ref{zestpt})--(\ref{zetupt}), we find
\begin{eqnarray}
        \beta_T &=& T \left( \zeta_{\widetilde D} - \zeta_D + \zeta_\lambda
                - \zeta_{\widetilde \lambda} \right) =
 \label{betatm} \\
        &=& - T (1 - T) {{\widetilde f} \over 2} \left[ {1 \over T} 
        + (n-1) \, {1 + 7w + 4 w^2 \over (1 + w)^3} \right] \ . \nonumber 
\end{eqnarray}
Clearly, the only possible fixed points here are $T^* = 1$, $T^* = 0$, and
$T^* = \infty$.
At the equilibrium fixed point $T^* = 1$, one finds 
$\zeta_\lambda = - (n-1) f / (1+w)$ and $\zeta_D = - f / 2$, and the beta 
functions for the couplings $w = \lambda / D$ and $f = g^2 / \lambda D$ read 
\cite{sssmod,sssfth}
\begin{eqnarray}
        \beta_w &=& w \left( \zeta_\lambda - \zeta_D \right) 
                = - w f \left( {n-1 \over 1 + w} - {1 \over 2} \right) \ ,
 \label{betweq} \\
        \beta_f &=& f \left( 2 \zeta_g - \zeta_\lambda - \zeta_D \right) =
                \nonumber \\
        &=& f \left[ - \epsilon + f \left( {n-1 \over 1 + w} + {1 \over 2} 
                \right) \right] \ .
 \label{betfeq}
\end{eqnarray}
The IR-stable fixed point (to one-loop order at least) turns out to be the 
strong-scaling SSS fixed point 
\begin{equation}
        w^* = 2n-3 \; , \quad f^* = \epsilon + O(\epsilon^2)
 \label{ssfpeq}
\end{equation}
with {\em equal} time scales governing the critical slowing down of the order 
parameter ($t^c_S \propto \xi^{z_S} \propto |\tau|^{- z_S \nu}$) and conserved 
generalized angular momenta fluctuations 
($t^c_M \propto \xi^{z_M} \propto |\tau|^{- z_M \nu}$), respectively,
\begin{equation}
        z_S = 2 + \zeta_\lambda^* \; , \quad
        z_M = 2 + \zeta_D^* \ .
 \label{cexzsm}
\end{equation}
Inserting the fixed-point values (\ref{ssfpeq}) yields the dynamic critical 
exponent
\begin{equation}
        z = z_S = z_M = 2 - {\epsilon \over 2} = {d \over 2} \ ,
 \label{cexzeq}
\end{equation}
which is actually an {\em exact} result, provided $z_S = z_M$ and 
$0 < f^* < \infty$ is finite, because Eq.~(\ref{betfeq}) then requires that
$2 z = 4 + \zeta_\lambda^* + \zeta_D^* = 4 + 2 \zeta_g^* = 4 - \epsilon = d$
\cite{modele,sssfth}.
The equilibrium static critical behavior is described by the zero of the beta 
function 
\begin{equation}
        \beta_u = u \zeta_u = u \left( - \epsilon + {n+8 \over 6} \, u \right)
                \ ,
 \label{hsfpeq}
\end{equation}
which yields of course the $O(n)$ Heisenberg fixed point
\begin{equation}
        u^* = {6 \over n+8} \, \epsilon + O(\epsilon^2)
 \label{heisfp}
\end{equation}
with the associated two independent critical exponents
\begin{eqnarray}
        &&\nu^{-1} = - \zeta_\tau^* = 2 - {n+2 \over n+8} \, \epsilon 
                + O(\epsilon^2) \ ,  
 \label{heisnu} \\
        &&\eta = - \zeta^* = 0 + O(\epsilon^2) \ .
 \label{heiset}
\end{eqnarray}

For the isotropic non-equilibrium fixed point with temperature ratio 
$T^* = \infty$, the appropriate effective mode-coupling constant becomes
\begin{equation}
        \overline{f} = {T \over w} \, {\widetilde f} 
                = {{\widetilde D} \over \lambda} \, f 
                = {{\widetilde D} \over \lambda} \, {g^2 \over \lambda D} \ ,
 \label{tinmcp}
\end{equation}
in terms of which the beta functions read \cite{uwezol}
\begin{eqnarray}
        \beta_w &=& w \left( \zeta_\lambda - \zeta_D \right) 
                = - (n-1) \, {w^4 \, \overline{f} \over (1 + w)^3} \ ,
 \label{btwne1} \\
        \beta_{\overline{f}} &=& \overline{f} \left( 2 \zeta_g +
        \zeta_{\widetilde D} - 2 \zeta_\lambda - \zeta_D \right) = \nonumber \\
        &=& \overline{f} \left( - \epsilon + 2 (n-1) \, 
                {w^3 \, \overline{f} \over (1 + w)^3} \right) \ ,
 \label{btfne1} \\
        \beta_{\widetilde u} &=& {\widetilde u} \left( \zeta_u 
                + \zeta_{\widetilde \lambda} - \zeta_\lambda \right) = 
 \label{btune1} \\
        &=& {\widetilde u} \left( - \epsilon + {n+8 \over 6} \, {\widetilde u}
        - 2 (n-1) \, {w (1 + 3w + w^2) \over (1 + w)^3} \, \overline{f} \right)
                \ . \nonumber
\end{eqnarray}
Thus, the RG fixed points to $O(\epsilon)$ governing this scaling regime are
\begin{equation}
        w^* = \infty \; , \quad \overline{f}^* = {\epsilon \over 2 (n-1)} \; ,
        \quad {\widetilde u}^* = {12 \over n+8} \, \epsilon \ .
 \label{sfpne1}
\end{equation}
Notice that the fixed point of the static coupling takes on twice the 
Heisenberg value (\ref{heisfp}); correspondingly, the ``static'' exponents will
be modified as compared to the equilibrium results.
E.g., the correlation length exponent now becomes
\begin{equation}
        \nu^{-1} = 2 - 2 \, {n+2 \over n+8} \, \epsilon - {\epsilon \over 2} 
                + O(\epsilon^2)
 \label{nuxne1}
\end{equation}
instead of Eq.~(\ref{heiset}), while to one-loop order both the order parameter
response and correlation function are characterized by the Wilson-Fisher
exponent $\eta = 0 + O(\epsilon^2)$ as in equilibrium \cite{uwezol}.
The characteristic time scales for the order parameter and the conserved
quantities are now governed by different power laws, namely
\begin{equation}
        z_S = 2 - {\epsilon \over 2} = {d \over 2} \; , \quad z_M = 2 \ .
 \label{exzne1}
\end{equation}
As $T^* = \infty$ means that effectively the heat bath for the conserved
generalized angular momenta is at infinite temperature, there is effectively an
energy current into the order parameter heat bath, but no feedback.
This explains why we find the coupled SSS model dynamic exponent for the order 
parameter fluctuations, while the generalized angular momenta correlations
decay {\em faster} with the purely diffusive exponent $z_M = 2$.
Finally, at the critical point there are non-trivial noise correlations
$\propto q^\rho$ only for the order parameter noise,
\begin{equation}
        \rho_S = 2 + \zeta_{\widetilde \lambda}^* = 2 - {3 \over 2} \, \epsilon
                + O(\epsilon^2) \ ,
 \label{ncsne1}
\end{equation}
while for the generalized angular momenta
\begin{equation}
        \rho_M = 2 + \zeta_{\widetilde D}^* = 2 \ ,
 \label{ncmne1}
\end{equation}
as to be expected at infinite temperature \cite{uwezol}.

For the other isotropic non-equilibrium fixed point, being characterized by 
$T^* = 0$, the correct effective mode-coupling constant reads
\begin{equation}
        {\widetilde f}' = {{\widetilde f} \over T}
        = {{\widetilde \lambda^2} D \over \lambda^2 {\widetilde D}} \, 
                {g^2 \over \lambda D} 
 \label{tzemcp}
\end{equation}
(called ${\tilde f}$ in Ref.~\cite{uwezol}), and consequently $\beta_w = 0$, 
which leaves the fixed point $w^*$ undetermined.
The remaining RG beta functions are \cite{uwezol}
\begin{eqnarray}
        \beta_{{\widetilde f}'} &=& {\widetilde f}' \left( 2 \zeta_g +
                2 \zeta_{\widetilde \lambda} - 3 \zeta_\lambda 
                - \zeta_{\widetilde D} \right) = {\widetilde f}' \left( 
        - \epsilon + {1 \over 2} \, {\widetilde f}' \right) \, ,\nonumber \\ &&
 \label{btfne2} \\
        \beta_{\widetilde u} &=& {\widetilde u} \left( \zeta_u 
                + \zeta_{\widetilde \lambda} - \zeta_\lambda \right) 
        = {\widetilde u} \left( - \epsilon + {n+8 \over 6} \, {\widetilde u}
        \right) \, ,
 \label{btune2}
\end{eqnarray}
with the $O(\epsilon)$ fixed points
\begin{equation}
        {\widetilde f'}\phantom{}^* = 2 \epsilon + O(\epsilon^2) \; , \quad 
        {\widetilde u}^* = {6 \over n+8} \, \epsilon + O(\epsilon^2) \ .
 \label{sfpne2}
\end{equation}
As ${\widetilde u}^*$ is identical to the Heisenberg fixed point 
(\ref{heisfp}), it turns out that the static exponents are indeed those of the
equilibrium static theory, (\ref{heisnu}) and (\ref{heiset}), both for the
order parameter response and correlation functions.
Now the energy current flows from the order parameter heat bath towards the
conserved generalized angular momenta, and thus the mode-coupling effects 
become negligible for the order parameter fluctuations.
The dynamic exponents therefore become model-A like, with
\begin{equation}
        z_S = z_M = 2 
 \label{exzne2}
\end{equation}
to one-loop order, and with the critical noise exponents
\begin{equation}
        \rho_S = 2 \; , \quad \rho_M = 2 - \epsilon = d - 2 \ .
 \label{ncrne2}
\end{equation}
Again, we observe that these anomalous power laws for the noise correlators
apply for those quantities which are governed by the heat bath at {\em lower}
temperature; here, the generalized angular momenta at effectively $T = 0$
\cite{uwezol}.

The above discussion of the isotropic fixed points facilitates the 
interpretation of the novel fixed points of the SSS model with dynamic 
anisotropy.
From Eq.~(\ref{betsig}) we had already inferred that at the fixed point with
$\sigma^* = 0$, one must have ${T^\parallel}^* = \infty$ and ${T^\perp}^* = 0$.
In analogy with Eqs.~(\ref{tinmcp}) and (\ref{tzemcp}), it is therefore 
convenient to introduce new effective mode-coupling constants in the
longitudinal and transverse sectors, respectively,
\begin{eqnarray}
        \overline{f}^\parallel &=& 
                {T^\parallel \over w^\parallel} \, {\widetilde f}^\parallel 
        = {{\widetilde D}^\parallel \over \lambda^\parallel} \, 
                {g^2 \over \lambda^\parallel D^\parallel} \ ,
 \label{atinmc} \\
        {\widetilde f'}\phantom{}^\perp &=& 
                {{\widetilde f}^\perp \over T^\perp}
        = {{\widetilde {\lambda^\perp}^2} D^\perp \over {\lambda^\perp}^2 
                {\widetilde D}^\perp} \, {g^2 \over \lambda^\perp D^\perp} \ .
 \label{atzemc}        
\end{eqnarray}
In terms of these couplings, one finds for $\sigma^* = 0$
\begin{eqnarray}
        \beta_{T^\parallel} &=& {\overline{f}^\parallel \over 2} \Biggl[ 
                w^\parallel \left( 1 - {1 \over T^\parallel} \right) - 
 \label{bettpa} \\
        &&- (n-1) \left( 1 - {d_\parallel \over d} \, T^\parallel \right) 
                {w^\parallel (1 + 7 w^\parallel + 4 {w^\parallel}^2) \over 
                (1 + w^\parallel)^3} + \nonumber \\
        &&\qquad \qquad + {n-1 \over 3} \, {d_\perp \over d} \, T^\parallel
        {w^\parallel (1 + 3 w^\parallel) \over (1 + w^\parallel)^3} \Biggr] \ ,
                \nonumber \\
        \beta_{T^\perp} &=& - T^\perp {{\widetilde f'}\phantom{}^\perp \over 2}
                \Biggl[ 1 - T^\perp + 
 \label{betttr} \\
        &&+ (n-1) T^\perp \left( 1 - {d_\parallel \over d} \, T^\parallel 
        \right) {1 + 7 w^\perp + 4 {w^\perp}^2) \over (1 + w^\perp)^3} + 
                \nonumber \\
        &&\qquad \qquad + {n-1 \over 3} \, {d_\parallel \over d} \, T^\parallel
        \,T^\perp\, {1 + 3 w^\perp) \over (1 + w^\perp)^3} \Biggr]\ , \nonumber
\end{eqnarray}
which indeed lead to the expected fixed points
\begin{equation}
       {T^\parallel}^* = \infty \; , \quad {T^\perp}^* = 0 \ ,
 \label{asfptt}
\end{equation}
with ${T^\parallel}^* \, {T^\perp}^* = 0$.

In the longitudinal sector, we may then write the beta functions for
$w^\parallel$, $\overline{f}^\parallel$, and ${\widetilde u}^\parallel$ as
\begin{eqnarray}
        \beta_{w^\parallel} &=& - {n-1 \over 3} 
                \left( 1 + 2 {d_\parallel \over d} \right) {{w^\parallel}^4 \,
                \overline{f}^\parallel \over (1 + w^\parallel)^3}\ ,
 \label{betwpa} \\
        \beta_{\overline{f}^\parallel} &=& \overline{f}^\parallel \left[ -
        \epsilon + {2 (n-1) \over 3} \left( 1 + 2 {d_\parallel \over d} \right)
        {{w^\parallel}^3 \, \overline{f}^\parallel \over (1 + w^\parallel)^3} 
                \right] \ , \nonumber \\ &&
 \label{betfpa} \\
        \beta_{{\widetilde u}^\parallel} &=& {\widetilde u}^\parallel \Biggl[
        - \epsilon + {n+8 \over 6} \, {\widetilde u}^\parallel - \nonumber \\
        &&\qquad - 2 (n-1) \, {d_\parallel \over d} \, 
                {w^\parallel (1 + 3 w^\parallel + {w^\parallel}^2) \over 
                (1 + w^\parallel)^3} \, \overline{f}^\parallel + \nonumber \\
        &&\qquad + {2 (n-1) \over 3} \, {d_\perp \over d} \, {{w^\parallel}^3 
        \, \overline{f}^\parallel \over (1 + w^\parallel)^3} \Biggr] \ ,
 \label{betupa}
\end{eqnarray}
yielding the stable fixed point
\begin{equation}
        {w^\parallel}^* = \infty \ ,
 \label{asfpwp}
\end{equation}
and consequently
\begin{eqnarray}
      {\overline{f}^\parallel}^* &=& 
      {3 \over 2 (n-1) (1 + 2 d_\parallel / d)} \, \epsilon + O(\epsilon^2) \ ,
 \label{asfpfp} \\
      {\widetilde u}^\parallel\phantom{}^* &=& {36 \, d_\parallel / d \over 
              (n+8) (1 + 2 d_\parallel / d)} \, \epsilon + O(\epsilon^2) \ .
\end{eqnarray}
For $d_\perp = 0$, i.e., $d_\parallel = d$, these expressions reduce to the
isotropic fixed point (\ref{sfpne1}) with $T^* = \infty$, and in fact, we
arrive at very similar results for the dynamic exponents,
\begin{eqnarray}
        z_S^\parallel &=& 2 + \zeta_{\lambda^\parallel}^* = 
                2 - {\epsilon \over 2} = {d \over 2} \ ,
 \label{aszspa} \\
        z_M^\parallel &=& 2 + \zeta_{D^\parallel}^* = 2 \ ,
 \label{aszmpa}
\end{eqnarray}
as well as for the anomalous noise exponents,
\begin{eqnarray}
        \rho_S^\parallel &=& 2 + \zeta_{{\widetilde \lambda}^\parallel}^* = 
        2 - {1 + 8 d_\parallel / d \over 2 (1 + 2 d_\parallel / d)} \, \epsilon
                + O(\epsilon^2) \ ,
 \label{asnspa} \\
        \rho_M^\parallel &=& 2 + \zeta_{{\widetilde D}^\parallel}^* = 2 \ ,
 \label{asnmpa}
\end{eqnarray}
compare Eqs.~(\ref{exzne1}) and (\ref{ncsne1}), (\ref{ncmne1}).
Moreover, we again find a non-standard correlation length exponent
\begin{eqnarray}
        {\nu^\parallel}^{-1} &=& - \zeta_{\tau^\parallel}^* 
        = 2 - {6 (n+2) d_\parallel / d \over (n+8) (1 + 2 d_\parallel / d)} \, 
                \epsilon + \nonumber \\
        &&\qquad + {1 - 4 d_\parallel / d \over 2 (1 + 2 d_\parallel / d)} \, 
                \epsilon + O(\epsilon^2) \ ,
 \label{asnupa}
\end{eqnarray}
but, in addition, non-trivial Wilson-Fisher exponents describing the critical
decay of the order parameter response and correlation functions, respectively,
\begin{eqnarray}
        \eta^\parallel &=& - {\zeta^\parallel}^* = {1 - d_\parallel / d 
               \over 2 (1 + 2 d_\parallel / d)} \, \epsilon + O(\epsilon^2) \ ,
 \label{asetpa} \\
        \eta_S^\parallel &=& - {\zeta_S^\parallel}^* = - {1 - d_\parallel / d 
        \over 2 (1 + 2 d_\parallel / d)} \, \epsilon + O(\epsilon^2) \ ;
 \label{asespa}
\end{eqnarray}
remarkably, $\eta_S^\parallel = - \eta^\parallel$ to $O(\epsilon)$.
These anomalous critical exponents at the anisotropic fixed point, appearing in
the longitudinal sector with effectively infinite heat bath temperature for the
generalized angular momenta, are obviously a consequence of the spatially 
extremely anisotropic noise correlations in the conserved quantities.
They may perhaps be interpreted as remnants of the elastic, pseudo-dipolar 
interactions generated in a model with conserved order parameter and dynamical 
anisotropy \cite{bearoy,schmit,kevzol}.

We now turn to the transverse sector, with the conserved quantities being 
effectively at zero temperature as ${T^\perp}^* = 0$.
Quite as at the isotropic non-equilibrium SSS model fixed point with $T^* = 0$,
we find that $\beta_{w^\perp} = 0$ and hence ${w^\perp}^*$ has no fixed value,\
while
\begin{eqnarray}
        \beta_{{\widetilde f'}\phantom{}^\perp} &=& 
                {\widetilde f'}\phantom{}^\perp \left( 
        - \epsilon + {1 \over 2} \, {\widetilde f'}\phantom{}^\perp \right) \ ,
 \label{betftr} \\
        \beta_{{\widetilde u}^\perp} &=& {\widetilde u}^\perp \left( 
                - \epsilon + {n+8 \over 6} \, {\widetilde u}^\perp \right) \ ,
 \label{betutr}
\end{eqnarray}
which yield the one-loop fixed points
\begin{equation}
        {{\widetilde f'}\phantom{}^\perp}^* = 2 \epsilon + O(\epsilon^2) \; , 
        \quad {{\widetilde u}^\perp}\phantom{}^* 
                = {6 \over n+8} \, \epsilon + O(\epsilon^2) \ .
 \label{asfptr}
\end{equation}
The ensuing critical exponents are precisely those of the $T^* = 0$ isotropic
non-equilibrium fixed point:
\begin{eqnarray}
        &&{\nu^\perp}^{-1} = - \zeta_{{\tau^\perp}^*} 
                = 2 - {n+2 \over n+8} \, \epsilon + O(\epsilon^2) \ ,  
 \label{asnutr} \\
        &&\eta^\perp = - {\zeta^\perp}^* = \eta_S^\perp = - {\zeta_S^\perp}^* 
                = 0 + O(\epsilon^2) \ ,
 \label{asettr}
\end{eqnarray}
and
\begin{eqnarray}
        &&z_S^\perp = 2 + \zeta_{\lambda^\perp}^* = 
                z_M^\perp = 2 + \zeta_{D^\perp}^* = 2 \ ,
 \label{aszsmt} \\
        &&\rho_S^\perp = 2 + \zeta_{{\widetilde \lambda}^\perp}^* = 2 \ ,
 \label{asnstr} \\
        &&\rho_M^\perp = 2 + \zeta_{{\widetilde D}^\perp}^* = 
                2 - \epsilon = d - 2 \ .
 \label{asnmtr}
\end{eqnarray}

We may interpret our results for the novel anisotropic fixed point as follows. 
For $T^\perp = 0$ and $T^\parallel = \infty$, the system breaks up into 
essentially independent sheets of dimension $d_\parallel$ with infinite heat
bath temperature.
The associated critical exponents are closely related to the isotropic ones
at the fixed point with $T^* = \infty$.
However, the additional $d_\perp$ dimensions are reflected in the anomalous
Wilson-Fisher exponents (\ref{asetpa}) and (\ref{asespa}), which are 
proportional to $d_\perp \, \epsilon$ (while the equilibrium exponent 
$\eta \propto \epsilon^2$).
Fluctuations in the $d_\perp$ transverse direction are effectively at zero
temperature for the conserved noise, and consequently are governed by the
critical exponents of the isotropic non-equilibrium SSS model with $T^* = 0$.
In the converse situation, with $T^\parallel = 0$ and $T^\perp = \infty$, 
clearly one must simply exchange the roles of the transverse and longitudinal
sectors.
Yet, we emphasize again that these novel fixed points with their associated
rather bizarre critical behavior are {\em unstable}, and for {\em any} 
$0 < \sigma = T^\perp / T^\parallel < \infty$ initially, the static and dynamic
critical properties of the system are {\em asymptotically} described by the 
{\em equilibrium} scaling exponents.

\section{The anisotropic non-equilibrium model J}
 \label{modelj}

In this section, we study the critical behavior of our non-equilibrium version
for model J (describing the dynamics of isotropic ferromagnets) with dynamical 
noise, as defined through Eqs.~(\ref{ferlan}) and (\ref{fernoi}).
We start by computing the $T_c$ shift from the static susceptibility.
As a consequence of the spatially anisotropic conserved noise with 
$T_0^\perp < T_0^\parallel$, it turns out that the transverse momentum space 
sector with {\em lower} noise temperature softens first.
Thus, at the critical point, the longitudinal sector remains uncritical 
(``stiff''), similar to equilibrium anisotropic elastic phase transitions 
\cite{elphtr}.
It is then instructive to switch off the mode-coupling constant $g_0$, and 
first recapitulate the properties of the ensuing two-temperature 
non-equilibrium model B \cite{bearoy,schmit,kevzol}.
In Sec.~\ref{mdjren}, we turn to the perturbational renormalization of
the two-temperature non-equilibrium model J to one-loop order, and finally
discuss the resulting RG flow equations.

\subsection{Dynamic field theory and the anisotropic ${\bf T_c}$ shift}
 \label{atcshf}

The probability distribution for the dynamic fields $S_0^\alpha$ 
($\alpha = 1,2,3$), equivalent to the Langevin equation (\ref{ferlan}) with
anisotropic noise correlator (\ref{fernoi}), reads
\begin{equation}
        P[\{ S_0^\alpha \}] \propto \int {\cal D}[\{ i {\widetilde S}_0^\alpha
        \}] \, e^{J[\{ {\widetilde S}_0^\alpha \}, \{ S_0^\alpha \}]} \ ,
 \label{genfnj}
\end{equation}
with the Janssen-De Dominicis functional 
$J = J_{\rm har} + J_{\rm rel} + J_{\rm mc}$, with the harmonic part
\begin{eqnarray}
        &&J_{\rm har}[\{ {\widetilde S}_0^\alpha \} , \{ S_0^\alpha \}] =
                \nonumber \\ 
        &&\quad = \int \! d^dx \int \! dt \sum_\alpha \Biggl\{ 
        -{\widetilde S}_0^\alpha \left( {\widetilde \lambda}_0^\parallel 
        \nabla_\parallel^2 + {\widetilde \lambda}_0^\perp \nabla_\perp^2 
                \right) {\widetilde S}_0^\alpha - \nonumber \\ 
        &&\qquad \qquad - {\widetilde S}_0^\alpha \left[ 
                {\partial \over \partial t} - \lambda_0 \nabla^2 
                \left( r_0 - \nabla^2 \right) \right] S_0^\alpha \Biggr\} \ ,
 \label{harfnj}
\end{eqnarray}
the non-linear relaxation vertex from model B,
\begin{equation}
        J_{\rm rel}[\{ {\widetilde S}_0^\alpha \} , \{ S_0^\alpha \}] =
        \lambda_0 {u_0 \over 6} \int d^dx \! \int dt \sum_{\alpha,\beta}
        {\widetilde S}_0^\alpha \nabla^2 S_0^\alpha S_0^\beta S_0^\beta \ , 
 \label{relfnj}
\end{equation}
and the mode-coupling vertices from the spin precession forces,
\begin{equation}
        J_{\rm mc}[\{ {\widetilde S}_0^\alpha \},\{ S_0^\alpha \}] = 
        - g_0 \int \! d^dx \int \! dt \! \sum_{\alpha,\beta,\gamma} 
                \epsilon^{\alpha \beta \gamma}
                {\widetilde S}_0^\alpha S_0^\beta \nabla^2 S_0^\gamma \ .
 \label{mctfnj}
\end{equation}

In analogy with Eq.~(\ref{opvsus}) for the SSS model, the dynamic 
susceptibility can be expressed as 
\begin{eqnarray}
        \chi_0({\bf q},\omega) &=&  
        \Gamma_{0\, {\widetilde S}S}(-{\bf q},-\omega)^{-1} \times \nonumber \\
        &&\times \left[ \lambda_0 \, q^2 + g_0 \,
        \Gamma_{0\, {\widetilde S}[{\widetilde S}S]}(-{\bf q},-\omega) 
        \right] \ .
 \label{opjsus}
\end{eqnarray}
From the ensuing expression (to one-loop order), we may determine the 
fluctuation-induced shift of the critical temperature.
Because of the dynamic anisotropy appearing in the noise correlator 
(\ref{fernoi}), however, the result depends on how the limit 
${\bf q} \to {\bf 0}$ is taken; upon defining $q_\parallel = q \cos \Theta$ and
$q_\perp = q \sin \Theta$, we find
\begin{eqnarray}
        &&r_{0c}(\Theta) = - \left( {d_\parallel \over d} \, T_0^\parallel 
                + {d_\perp \over d} \, T_0^\perp \right) \times
 \label{tcjint} \\
        &&\quad \times \left[ {5 \over 6} \, u_0 \int_k {1 \over r_{0c} + k^2}
        - {g_0^2 \over 2 (d+2) \lambda_0^2} \int_k {1 \over k^2(r_{0c} + k^2)}
                \right] - \nonumber \\
        &&-\left( T_0^\parallel \cos^2 \Theta + T_0^\perp \sin^2 \Theta \right)
        {g_0^2 \over 2 (d+2) \, \lambda_0^2} \int_k {1 \over k^2(r_{0c} + k^2)}
        \ , \nonumber
\end{eqnarray}
in contrast with the isotropic Eq.~(\ref{tcsint}) for the SSS model.
As $T_c = T_c^0 + r_{0c}$, the phase transition will occur at the maximum of 
the function $r_{0c}(\Theta)$, which for $T_0^\perp < T_0^\parallel$ occurs at
$\Theta = \pi / 2$.
The $d_\perp$-dimensional transverse sector in momentum space thus softens 
first, and the true $T_c$ shift is given by
\begin{eqnarray}        
        &&r_{0c} = - \left( {d_\parallel \over d} \, T_0^\parallel 
                + {d_\perp \over d} \, T_0^\perp \right) {5 \over 6} \, u_0 
                \int_k {1 \over r_{0c} + k^2} + \nonumber \\
        &&\; + {d_\parallel \over d} \left( T_0^\parallel - T_0^\perp \right) 
        {g_0^2 \over 2 (d+2) \, \lambda_0^2} \int_k {1 \over k^2(r_{0c} + k^2)}
        \ ,
 \label{tctint}
\end{eqnarray}
or, after evaluating the integrals in Eq.~(\ref{tctint}) by means of 
dimensional regularization, as solution of
\begin{eqnarray}
        |r_{0c}|^{6-d \over 2} &=& {2 A_d \over (d-2) (4-d)} \Biggl[
        \left( {d_\parallel \over d} \, T_0^\parallel + {d_\perp \over d} \,
        T_0^\perp \right) {5 \over 6} \, u_0 |r_{0c}| + \nonumber \\
        &&\qquad + {d_\parallel \over d} \left( T_0^\parallel - T_0^\perp 
                \right) {g_0^2 \over 2 (d+2) \, \lambda_0^2} \Biggr] \ .
 \label{tcsheq}
\end{eqnarray}
(We remark again, though, that a more physical way to compute this quantity
would be by means of cutoff regula\-rization.)
For $T_0^\parallel = T_0^\perp = T_0$, we recover the equilibrium result with a
rescaled coupling $T_0 u_0 = {\widetilde \lambda}_0 u_0 / \lambda_0$, as to be
expected (see Ref.~\cite{uwezol} and Eq.~(\ref{tcshif}) for the SSS model with 
$n=3$). 
Notice, however, that dynamical anisotropy ($T_0^\parallel \not= T_0^\perp$),
{\em combined} with the reversible mode-coupling terms, has a very drastic
effect here: It renders the system soft only in the momentum subspace with 
lower noise temperature.
This effect has a simple physical interpretation: The $T_c$ shift is due to
thermal fluctuations, which are reduced in the transverse sector 
($T_0^\perp < T_0^\parallel$), and therefore lead to a comparatively stronger 
downwards shift in the longitudinal sector.

In order to characterize the critical properties of our model, we may neglect
terms $\propto q_\parallel^4$ in the stiff momentum space sector, because
$\tau_0^\parallel = r_0 - r_{0c}(\Theta = 0)$ remains positive at the phase 
transition where $\tau_0^\perp = r_0 - r_{0c}(\Theta = \pi / 2)$ vanishes.
In analogy with the situation at anisotropic elastic structural phase 
transitions \cite{elphtr}, or with Lifshitz points in magnetic systems with
competing interactions \cite{lifsht}, we thus have to scale the soft and stiff 
wavevector components differently, $[q_\perp] = \mu$, whereas 
$[q_\parallel] = [q_\perp]^2 = \mu^2$.
Consequently, while $[{\widetilde \lambda}_0^\perp] = \mu^0$ if we choose
$[\omega] = \mu^4$, we find for the longitudinal scaling dimension
$[{\widetilde \lambda}_0^\parallel] = \mu^{-2}$, which implies that the
longitudinal noise strength becomes {\em irrelevant} under scale 
transformations.   
Allowing for distinct couplings in the different sectors, one finds in the same
manner that the ratios $[\lambda_0^\parallel / \lambda_0^\perp] = 
[\lambda_0^\parallel u_0^\parallel / \lambda_0^\perp u_0^\perp] = 
[g_0^\parallel / g_0^\perp] = \mu^{-2}$ all have negative scaling dimension.
Thus, for an investigation of the asympotic critical behavior, the longitudinal
parameters may be neglected as compared to their transverse counterparts, and 
can all be set to zero in the effective dynamic functional $J$.
Upon rescaling the fields according to $S_0^\alpha \to 
({\widetilde \lambda}_0^\perp/\lambda_0^\perp)^{1/2} S_0^\alpha$, 
${\widetilde S}_0^\alpha \to 
(\lambda_0^\perp/{\widetilde \lambda}_0^\perp)^{1/2} {\widetilde S}_0^\alpha$,
defining
\begin{equation}
        c_0 = {\lambda_0^\parallel \over \lambda_0^\perp} \, \tau_0^\parallel
        \; , \quad 
        {\widetilde u}_0 = {{\widetilde \lambda}_0^\perp \over \lambda_0^\perp}
        \, u_0^\perp \; , \quad 
        {\widetilde g}_0 = \sqrt{{\widetilde \lambda}_0^\perp \over 
                \lambda_0^\perp} \, g_0^\perp \ ,
 \label{nejcpl}
\end{equation}
and omitting the labels ``$\perp$'' again for $\lambda_0$ and $r_0$, the 
ensuing {\em effective} Langevin equation of motion becomes
\begin{eqnarray}
        &&{\partial S_0^\alpha \over \partial t} =
        \lambda_0 \left[ c_0 \nabla_\parallel^2 + 
        \nabla_\perp^2 (r_0 - \nabla_\perp^2) \right] S_0^\alpha +
 \label{nejlan} \\
        &&\quad + \lambda_0 {{\widetilde u}_0 \over 6} \nabla_\perp^2 
                S_0^\alpha \sum_\beta S_0^\beta S_0^\beta       
        - {\widetilde g}_0 \sum_{\beta,\gamma} \epsilon^{\alpha \beta \gamma} 
        S_0^\beta \nabla_\perp^2 S_0^\gamma + \zeta^\alpha \ , \nonumber
\end{eqnarray}
with the noise correlator
\begin{equation}
        \langle \zeta^\alpha({\bf x},t) \zeta^\beta({\bf x}',t') \rangle =     
        - 2 \lambda_0 \nabla_\perp^2 \, \delta({\bf x} - {\bf x}') \, 
        \delta( t - t') \, \delta^{\alpha \beta} \ .
 \label{nejnoi}
\end{equation}
These equations define the {\em two-temperature non-equili\-bri\-um model J}.

It is interesting to note that the anisotropy of the $T_c$ shift in 
Eq.~(\ref{tcjint}) only occurs in the contribution $\propto g_0^2$.
In the non-equilibrium model B with dynamical anisotropy, the criticality 
condition for the response function remains isotropic, at least to one-loop 
order.
Thus, if one does not assume different critical temperatures in the purely 
diffusive non-linear Langevin equation to begin with, these are {\em not} 
generated, and one is not led to the two-temperature model B, which we shall
discuss below, as the correct effective theory for the phase transition.
Instead, the non-equilibrium perturbations appear to be {\em irrelevant} to 
this order, and the model is asymptotically described by the {\em equilibrium} 
critical exponents of model B, i.e., the static exponents of the $O(n)$
Heisenberg model, accompanied with the dynamic exponent $z = 4 - \eta$.

\subsection{The two-temperature non-equilibrium model B}
 \label{modelb}

Before we turn to the analysis of the two-temperature model J, derived above as
the {\em effective critical} theory for the non-equilibrium model J with 
dynamical anisotropy, we briefly summarize the results for the corresponding 
two-temperature non-equilibrium model B \cite{bearoy,schmit,kevzol}, which is 
defined by Eqs.~(\ref{nejlan}) and (\ref{nejnoi}) with vanishing mode-coupling 
term ${\widetilde g}_0 = 0$. 
We may thus generalize to arbitrary number of compenents $n$ again.

With this simplification, the resulting purely relaxational Langevin equation 
of motion can be written in the form
\begin{equation}
        {\partial S_0^\alpha \over \partial t} = \lambda_0 \nabla_\perp^2 \,
                {\delta H_{\rm eff}[ \{ S_0^\alpha \}] \over \delta S_0^\alpha}
                + \zeta^\alpha \ ,      
 \label{neblan}
\end{equation}
accompanied with the Gaussian noise (\ref{nejnoi}).
Notice that after the above rescaling, the Einstein relation between the 
diffusion constant and the noise strength is fulfilled; hence the 
two-temperature model B is effectively an {\em equilibrium} system, and 
describes diffusive relaxation into the stationary state with probability
distribution 
$P_{\rm eq}[\{ S_0^\alpha \}] \propto \exp (-H_{\rm eff}[\{ S_0^\alpha \}])$.
The effective free energy here,
\begin{eqnarray}
        &&H_{\rm eff}[\{ S_0^\alpha \}] = \nonumber \\
        &&\quad = {1 \over 2} \int \! {d^dq \over (2 \pi)^d} \sum_\alpha 
        {c_0 q_\parallel^2 + q_\perp^2 (r_0 + q_\perp^2) \over q_\perp^2} \, 
                S_0^\alpha({\bf q}) S_0^\alpha(-{\bf q}) + \nonumber \\
        &&\qquad \qquad + {{\widetilde u}_0 \over 4!} \int d^dx 
        \sum_{\alpha , \beta} S_0^\alpha({\bf x})^2 S_0^\beta({\bf x})^2 \ ,
 \label{effham}
\end{eqnarray}
contains long-range elastic interactions (uniaxial pseudo-dipolar for 
$d_\perp = 1$ \cite{fnote3}), as is evident from the harmonic part of 
Eq.~(\ref{effham}). 
These long-range, non-analytic interactions are generated by the dynamical
anisotropy in the original non-equilibrium model \cite{bearoy,schmit,kevzol}.
As critical fluctuations are now restricted to the $d_\perp$-dimensional
subsector, one expects that the upper critical dimension of this model is
reduced to
\begin{equation}
        d_c^{\rm st} = 4 - d_\parallel \ ,
 \label{stcrdm}
\end{equation}
which is confirmed through a direct scaling analysis of the free energy 
(\ref{effham}).
E.g., for a one-dimensional hard sector (uniaxial system), the critical 
dimension above which mean-field exponents become exact, is $d_c^{\rm st} = 3$.

In order to compute the scaling exponents below the upper critical dimension,
we have to renormalize the theory in the UV.
To this end, we introduce renormalized fields and parameters according to
\begin{eqnarray}
        &&{\widetilde S} = Z_{\widetilde S}^{1/2} \, {\widetilde S}_0 \; , 
        \quad S = Z_S^{1/2} \, S_0 \ ,
 \label{jfdren} \\
        &&\lambda = Z_\lambda \, \lambda_0 \; , \qquad c = Z_c \, c_0 \ ,
 \label{jlcren} \\
        &&\tau = Z_\tau \, \tau_0 \, \mu^{-2} \; , \; 
        {\widetilde u} = Z_{\widetilde u} \, {\widetilde u}_0 \, 
                A(d_\parallel,d_\perp) \, \mu^{-\epsilon} \ ,
 \label{jcpren}
\end{eqnarray}
where $\tau_0 = r_0 - r_{0c}$ as usual, $\epsilon = d_c^{\rm st} - d = 
4 - d - d_\parallel = 4 - 2 d_\parallel - d_\perp$ denotes the deviation from 
the upper critical dimension (\ref{stcrdm}), and we define the anisotropic 
geometric factor
\begin{equation}
        A(d_\parallel,d_\perp) = {\Gamma(3-d/2-d_\parallel/2) \, \Gamma(d/2) 
        \over c_0^{d_\parallel/2} \, 2^{d-1} \pi^{d/2} \, \Gamma(d_\perp/2)}
 \label{adpatr}
\end{equation}
with $A(0,d) = A_d$.

Yet, these renormalization constants are not entirely independent of each 
other.
First, as the two-temperature model B in the critical region is equivalent to 
an equilibrium system, there exists a fluctuation-dissipation theorem that 
connects the imaginary part of the dynamic susceptibility 
$\chi_0({\bf q},\omega)$ with the Fourier transform of the order parameter 
correlation function $C_0({\bf x},t;{\bf x}',t') \, \delta^{\alpha\beta} = 
\langle S_0^\alpha({\bf x},t) \, S_0^\beta({\bf x}',t') \rangle$,
\begin{equation}
        C_0({\bf q},\omega) = 
                {2 \over \omega} \; {\rm Im} \, \chi_0({\bf q},\omega)
 \label{fdtcor} \ .
\end{equation}
In terms of the two-point vertex functions, $C_0({\bf q},\omega) = 
- \Gamma_{0\, {\widetilde S}{\widetilde S}}({\bf q},\omega) / 
|\Gamma_{0\, {\widetilde S}S}({\bf q},\omega)|^2$, while for $g_0 = 0$ 
Eq.~(\ref{opjsus}) reduces to $\chi_0({\bf q},\omega) = \lambda_0 q_\perp^2 /
\Gamma_{0\, {\widetilde S}S}(-{\bf q},-\omega)$; thus the 
fluctu\-ation-dissipation theorem can equivalently be written as
\begin{equation}
        \Gamma_{0\, {\widetilde S}{\widetilde S}}({\bf q},\omega) = 
        {2 \lambda_0 \, q_\perp^2 \over \omega} \;
                {\rm Im} \, \Gamma_{0\, {\widetilde S}S}({\bf q},\omega)
 \label{fdtver} \ .
\end{equation}
Precisely the same relation must hold for the corresponding renormalized vertex
function, which implies the identity
\begin{equation}
        Z_\lambda \equiv \left( Z_S / Z_{\widetilde S} \right)^{1/2} \ .
 \label{mdbid1}
\end{equation}
Second, the equation of motion (\ref{neblan}) implies that the non-linear
relaxation vertices are proportional to the external momentum $q_\perp^2$, and
hence the loop contributions to $\Gamma_{0\, {\widetilde S}S}({\bf q},\omega)$
must vanish in the limit ${\bf q}_\perp \to {\bf 0}$.
Thus, to {\em all} orders in perturbation theory,
\begin{equation}
        \Gamma_{0\, {\widetilde S} S}({\bf q}_\parallel,{\bf q}_\perp={\bf 0},
                \omega) \equiv i \omega + \lambda_0 \, c_0 \, q_\parallel^2 \ .
 \label{gmstsi}
\end{equation}
This leads to the additional set of identities
\begin{equation}
        Z_{\widetilde S} \, Z_S \equiv 1 \; , \quad
        Z_\lambda \, Z_c \equiv 1 \ ,
 \label{mdbid2}
\end{equation}
and thus, using Eq.~(\ref{mdbid1}),
\begin{equation}
        Z_\lambda \equiv Z_c^{-1} \equiv Z_S \ .
 \label{mdbid3}
\end{equation}
At last, because of the absence of the composite-operator vertex function in 
the relation between the dynamic susceptibility and the two-point vertex 
function, we have 
\begin{equation}
        \chi({\bf q},\omega) = Z \, \chi_0({\bf q},\omega)
 \label{rnsusc} 
\end{equation}
with
\begin{equation}
        Z \equiv Z_S \ .
 \label{mdbid4}
\end{equation}

As in Sec.~\ref{sssflo} for the non-equilibrium SSS model, we can now define 
Wilson's zeta functions via logarithmic derivatives of the $Z$ factors with 
respect to the renormalization scale $\mu$, 
\begin{eqnarray}
        \zeta_{\widetilde S} &=& \mu {\partial \over \partial \mu} \bigg\vert_0
                \ln Z_{\widetilde S} \ ,
 \label{ztjflt} \\
        \zeta_S &=& \mu {\partial \over \partial \mu} \bigg\vert_0 \ln Z_S \ , 
 \label{ztjfld} \\
        \zeta_a &=& \mu {\partial \over \partial \mu} \bigg\vert_0 
                \ln {a \over a_0} \ ,
 \label{zjtpar}
\end{eqnarray}
where $\{ a\} = \lambda, c, \tau, u$, and write down the RG equations for
the vertex and response functions,
\begin{eqnarray}
        &&\Gamma_{{\widetilde S}^r S^l} \left( \mu, \{ {\bf q}_\parallel,
        {\bf q}_\perp,\omega \};\{ a \} \right) = \nonumber \\
        &&\quad = \exp \left\{ {1 \over 2} \int_1^\ell 
        \Bigl[ r \, \zeta_{\widetilde S}(\ell') + s \, \zeta_S(\ell') \Bigr] 
                {d \ell' \over \ell'} \right\} \times \nonumber \\ 
        &&\qquad \qquad \times \Gamma_{{\widetilde S}^r S^s} \left( \mu \ell, 
        \{ {\bf q}_\parallel,{\bf q}_\perp,\omega \};\{ a(\ell) \} \right) \ ,
 \label{sljcsy} \\
        &&\chi \left( \mu, \{ {\bf q}_\parallel,{\bf q}_\perp,\omega \};\{ a \}
                \right) = \exp\left\{ -\int_1^\ell \zeta_S(\ell')
                {d \ell' \over \ell'} \right\} \times \nonumber \\
        &&\qquad \times \, \chi \left( \mu \ell, \{ {\bf q}_\parallel,
                {\bf q}_\perp,\omega \};\{ a(\ell) \} \right) \ . 
 \label{sljcsc}
\end{eqnarray}
The general scaling form for the renormalized order parameter response and 
correlation function thus obtained at an IR-stable fixed point 
${\widetilde u}^*$ reads \cite{schmit}
\begin{eqnarray}
        \chi(\tau,{\bf q}_\parallel,{\bf q}_\perp,\omega) &=& 
        q_\perp^{- 2 + \eta} \, {\hat \chi}\left( {\tau \over q_\perp^{1/\nu}},
        {q_\parallel \over q_\perp^{1+\Delta}},{\omega \over q_\perp^z}\right)
        \ , 
 \label{resscm} \\
        C(\tau,{\bf q}_\parallel,{\bf q}_\perp,\omega) &=&
        q_\perp^{- 2 - z + \eta}\, {\hat C}\left( {\tau \over q_\perp^{1/\nu}},
        {q_\parallel \over q_\perp^{1+\Delta}},{\omega \over q_\perp^z}\right)
        \ , \nonumber \\ &&
 \label{corscm}
\end{eqnarray}
where in addition to the usual static exponents $\eta$ and $\nu$, and the
dynamic exponent $z$, we have introduced a scaling exponent $\Delta$
originating in the intrinsic anisotro\-py of the system.
Alternatively, we could have defined a set different transverse and 
longitudinal critical exponents, $\nu_\perp = \nu$, $z_\perp = z$,
$\nu_\parallel = \nu / (1 + \Delta)$, $z_\parallel = z / (1 + \Delta)$
\cite{schmit}.
Notice that as a consequence of the identity (\ref{mdbid4}), the exponent 
$\eta$ governs the critical decay of the response as well as the correlation 
function, as is required by the fluctuation-dissipation theorem (\ref{fdtcor}).
The critical exponents are readily identified with the fixed-point values of
the flow functions
\begin{eqnarray}
        &&\eta = - \zeta_S^* \; , \qquad \nu^{-1} = - \zeta_\tau^* \ , 
 \label{mdjstx} \\
        &&z = 4 + \zeta_\lambda^* \; ,\quad \Delta = 1 - {\zeta_c^* \over 2}\ .
 \label{mdjdne}
\end{eqnarray}
The exact relation $Z_\lambda \equiv Z_c^{-1}$ implies 
$\zeta_\lambda \equiv - \zeta_c$, and consequently
\begin{equation}
        1 + \Delta \equiv z / 2 \ ,
 \label{delzid}
\end{equation}
or $z_\parallel \equiv 2$, which reflects the mean-field character of the 
fluctuations in the stiff sector, and obviously holds whenever 
Eq.~(\ref{gmstsi}) is valid.
From the second equation in (\ref{mdbid3}) we furthermore infer 
$\zeta_\lambda \equiv \zeta_S$ for the two-temperature model B, and hence
\cite{schmit}
\begin{equation}
        z \equiv 4 - \eta \; , \quad \Delta \equiv 1 - \eta / 2 \ .
 \label{etazid}
\end{equation}

The task is thus to compute the remaining independent static exponents $\eta$
and $\nu$.
To one-loop order, from ${1 \over 2} (\partial_{q_\perp^2} )^2 \,
\Gamma_{{\widetilde S} S}({\bf q}_\parallel = {\bf 0},{\bf q}_\perp,\omega = 0)
\vert_{{\bf q}_\perp={\bf 0}}$ one finds $Z_\lambda \equiv Z_S = 1$.
Next, the criticality condition for the static susceptibility yields the $T_c$ 
shift 
\begin{equation}
        |r_{0c}| = \left[ {n+2 \over 3} \, {{\widetilde u}_0 \, 
        A(d_\parallel,d_\perp) \over (d+d_\parallel-2) \, (4-d-d_\parallel)} 
                \right]^{2 \over 4 - d - d_\parallel} \ ;
 \label{mdbtcs}
\end{equation}
the denominator here indicates that in addition to the reduction of the upper 
critical dimension $d_c^{\rm st}$, the {\em lower} critical dimension appears 
to be lowered by $d_\parallel$ as well to $d_{\rm lc} = 2 - d_\parallel$.
Thus, in two dimensions, an ordered phase with long-range order may exist at 
low temperatures and is indeed found \cite{kevzol}; notice that the 
Mermin-Wagner-Hohenberg theorem is invalidated by the existence of 
{\em long-range} elastic or pseudo-dipolar interactions in the system.
Upon then replacing $r_0 = \tau_0 + r_{0c}$, rendering $\partial_{q_\perp^2} \,
\Gamma_{{\widetilde S} S}({\bf q}_\parallel = {\bf 0},{\bf q}_\perp,\omega = 0)
\vert_{{\bf q}_\perp={\bf 0}}$ UV-finite gives
\begin{equation}
        Z_\tau = 1 - {n+2 \over 6 \, \epsilon} \, {\widetilde u}_0 \, 
                A(d_\parallel,d_\perp) \, \mu^{-\epsilon} \ ,
 \label{mdbztt}
\end{equation}
while the four-point vertex function $\Gamma_{0 \, {\widetilde S} S S S}$
provides us with
\begin{equation}
        Z_{\widetilde u} = 1 - {n+8 \over 6 \, \epsilon} \, {\widetilde u}_0 \,
                A(d_\parallel,d_\perp) \, \mu^{-\epsilon} \ .
 \label{mdbztu}
\end{equation}
Notice that the combinatorics for the Feynman graphs of the anisotropic 
two-temperature model B is identical to the equilibrium model B, and therefore
the above renormalization constants assume the same form as their familiar 
equilibrium counterparts, apart from the shifted critical dimension and a
modified geometric factor $A(d_\parallel,d_\perp)$.
The one-loop RG flow functions thus are
\begin{eqnarray}
        &&\zeta_\lambda \equiv - \zeta_c \equiv \zeta_S = 0 \ ,
 \label{mdbfls} \\
        &&\zeta_\tau = - 2 + {n+2 \over 6} \, {\widetilde u} \,
 \label{mdbflt} \\
        &&\beta_{\widetilde u} = {\widetilde u} \, \zeta_{\widetilde u} 
        = {\widetilde u} \, \left( - \epsilon + {n+8 \over 6} \, {\widetilde u}
                \right) \ ,
 \label{mdbflu}
\end{eqnarray}
with the stable fixed point [compare Eq.~(\ref{heisfp})]
\begin{equation}
        {\widetilde u}^* = {6 \over n+8} \, \epsilon + O(\epsilon^2) \ ,
 \label{mdbfpu}
\end{equation}
leading to the $O(\epsilon)$ exponents, with $\epsilon = 4 - d - d_\parallel$,
\begin{equation}
        \eta = O(\epsilon^2) \; , \quad 
        \nu^{-1} = 2 - {n+2 \over n+8} \, \epsilon + O(\epsilon^2) \ .
 \label{mdbcex}
\end{equation}
For the two-loop fixed point and exponent values to $O(\epsilon^2)$ for the
case $n = 1$, we refer to Refs.~\cite{schmit,kevzol}.

\subsection{Renormalization of the two-temperature model J}
 \label{mdjren}

We now return to the two-temperature model J with non-vanishing mode-coupling
term, as defined in Eqs.~(\ref{nejlan}) and (\ref{nejnoi}).
In equilibrium dynamics, the reversible spin precession force of model J 
constitutes a {\em relevant} perturbation to the purely diffusive model B with
conserved three-component order parameter, and the ensuing non-trivial fixed
point (with upper critical dimension $d_c^{\rm dy} = 6$) for the renormalized
effective mode-coupling constant $f \propto g^2 / \lambda^2$ changes the 
dynamic critical exponent from $z \equiv 4 - \eta $ to 
$z \equiv (d + 2 - \eta) / 2$, describing {\em faster} relaxation processes
\cite{modelj,dynfun,erwrev}.
As we should expect the reversible mode coupling to be relevant in the
two-temperature variant of model J as well, the issue therefore is, does the
RG flow again lead to a non-trivial stable fixed point, and what are the values
of the ensuing critical exponents?

It is essential to note, however, that the two-tempera\-ture model J with
${\widetilde g}_0 \not= 0$ {\em cannot} be recast as the dynamics of an
equivalent equilibrium model, with an effective free energy (\ref{effham}),
because for the reversible forces in Eq.~(\ref{nejlan}),
\begin{equation}
        \int \! d^dx \, {\delta \over \delta S_0^\alpha(x)} \left( 
        - {\widetilde g}_0 \sum_{\beta,\gamma} \epsilon^{\alpha \beta \gamma} 
                S_0^\beta \nabla_\perp^2 S_0^\gamma \, 
                e^{- H_{\rm eff}} \right) \not= 0 \ ,
 \label{intgcd}
\end{equation}
and the necessary Deker--Haake integrability condition \cite{intcon} that would
ensure the stability of the equilibrium probability distribution 
$\propto \exp (- H_{\rm eff}[\{ S_0^\alpha \}])$ is violated, except for 
$c_0 = 0$ or $d_\parallel = 0$.
We remark that this is actually a consequence of the inseparability of 
``statics'' and dynamics in the intrinsically dynamic two-temperature model 
{\em with} reversible mode couplings.
For, if we could {\em first} and separately consider the effective {\em static}
free energy, and only {\em subsequently} introduce the dynamics with the analog
of Eq.~(\ref{ferlao}), the elastic pseudo-dipolar propagator would appear in
the mode-coupling vertex, and the above integrability condition
\begin{equation}
        \int \! d^dx \, {\delta \over \delta S_0^\alpha(x)} \left( 
        - {\widetilde g}_0 \sum_{\beta,\gamma} \epsilon^{\alpha \beta \gamma} 
                \, {\delta H_{\rm eff} \over \delta S_0^\beta} \, S_0^\gamma \,
                e^{- H_{\rm eff}} \right) = 0 
 \label{fintcd}
\end{equation}
would be satisfied.
Following the standard equilibrium procedures \cite{dynfun}, the ensuing
critical exponents would be given by Eqs.~(\ref{mdbcex}) and (\ref{delzid}), 
with the exact dynamic exponent $z = (d + 2 - \eta) / 2$.
Yet, such a procedure is not possible here, and the derivation of the effective
equation of motion has to proceed in the {\em dynamic} functional, leading to 
Eq.~(\ref{nejlan}).
As opposed to the two-temperature model B, the two-temperature model J thus
represents a {\em genuinely non-equilibrium} dynamical model, for, as we shall
see below, the renormalization-group flow does {\em not} take the renormalized
mode coupling ${\widetilde g}$ to zero under scale transformations.

Thus we can invoke no fluctuation-dissipation theorem in order to relate vertex
and response function renormalizations, and we have to compute almost all the 
$Z$ factors, as defined in Eqs.~(\ref{jfdren})--(\ref{jcpren}) and 
(\ref{rnsusc}), independently.
Neither is there a Ward identity relating the renormalization of the 
mode-coupling vertex to simple field renormalizations \cite{fnote4}.
Fortunately, though, because of the momentum dependence of the mode-coupling 
vertex, $\Gamma_{0 \,{\widetilde S}^\alpha S^\beta S^\gamma}(-{\bf q}_\perp,0;
{{\bf q}_\perp \over 2}-{\bf p}_\perp,0;$
${{\bf q}_\perp \over 2}+{\bf p}_\perp,0) = {\widetilde g}_0 \, ({\bf q}_\perp
\cdot {\bf p}_\perp) \, \epsilon^{\alpha\beta\gamma} + O({\widetilde g}_0^3)$, 
at least Eq.~(\ref{gmstsi}) is valid for the two-temperature model J as well.
Consequently the identities (\ref{mdbid2}) still hold, leading immediately to 
Eq.~(\ref{delzid}) or $z_\parallel \equiv 2$, as to be expected.
Furthermore, simple power counting yields that the dynamical upper critical
dimension, where the mode-coupling constant becomes dimensionless, is
\begin{equation}
        d_c^{\rm dy} = 6 - d_\parallel \ ,
 \label{dycrdm}
\end{equation}
i.e., the spatial anisotropy reduces $d_c^{\rm dy}$ from its equilibrium value
in exactly the same way by $d_\parallel$ as the static upper critical dimension
(\ref{stcrdm}).
Thus, we define a dimensionless renormalized mode-coupling constant according 
to
\begin{equation}
        {\widetilde g} = Z_{\widetilde g}^{1/2} \, {\widetilde g}_0 \, 
                B(d_\parallel,d_\perp)^{1/2} \, \mu^{- \varepsilon / 2} \ ,
 \label{jmcren}
\end{equation}
where $\varepsilon = d_c^{\rm dy} - d = 6 - d - d_\parallel = 
6 - 2 d_\parallel - d_\perp$, and 
\begin{equation}
        B(d_\parallel,d_\perp) = {\Gamma(4-d/2-d_\parallel/2) \, \Gamma(d/2) 
        \over c_0^{d_\parallel/2} \, 2^d \pi^{d/2} \, \Gamma(d_\perp/2)} \ .
 \label{bdpatr}
\end{equation}
The appearance of {\em two different} upper critical dimensions implies that we
shall have to compute fixed points and exponents in a {\em double} expansion
with respect to $\epsilon$ and $\varepsilon$.

In order to evaluate the renormalization constants, we start with the dynamic
susceptibility (\ref{opjsus}), and first compute the fluctuation-induced 
$T_c$ shift from the condition 
$\chi_0({\bf q}_\parallel={\bf 0},{\bf q}_\perp \to {\bf 0},\omega=0)^{-1} = 0$
for $r_0 = r_{0c}$.
Introducing the effective mode-coupling constant
\begin{equation}
        {\widetilde f}_0 = {{\widetilde g}_0^2 \over 2 d_\perp \, \lambda_0^2}
        \ ,
 \label{jefmcp}
\end{equation}
we find in dimensional regularization
\begin{eqnarray}
        &&|r_{0c}|^{{6-d-d_\parallel} \over 2} = {5 {\widetilde u}_0 \over 3} 
        \, {A(d_\parallel,d_\perp) \, |r_{0c}| \over (d+d_\parallel-2)
                (4-d-d_\parallel)} - \nonumber \\
        &&\qquad - {d_\parallel d_\perp \, {\widetilde f}_0 \over (d-2)} \, 
        {B(d_\parallel,d_\perp) \over (4-d-d_\parallel)\, (6-d-d_\parallel)}\ ,
 \label{mdjtcs}
\end{eqnarray}
to be compared with Eq.~(\ref{mdbtcs}) for the two-temperature model B.
Subsequently rendering both $\partial_{q_\perp^2} \, 
\chi({\bf q}_\parallel = {\bf 0},{\bf q}_\perp,$ $\omega = 0)^{-1} 
\vert_{{\bf q}_\perp={\bf 0}}$ and 
$\chi({\bf q}_\parallel = {\bf 0},{\bf q}_\perp \to {\bf 0},\omega = 0)$
UV-finite yields, after a somewhat tedious calculation, the $Z$ factors
\begin{eqnarray}
        Z &=& 1 - {d_\parallel (6-d_\parallel) \over 2 (4-d_\parallel)} \,
        {{\widetilde f}_0 \, B(d_\parallel,d_\perp) \, \mu^{-\varepsilon} \over
                \varepsilon} \ ,
 \label{mdjzet} \\
        Z \, Z_\tau &=& 1 - {5 \over 6} \, {{\widetilde u}_0 \, 
        A(d_\parallel,d_\perp) \, \mu^{-\epsilon} \over \epsilon} - \nonumber\\
        &&- {d_\parallel (3-d_\parallel) \over 4-d_\parallel} \, 
        {{\widetilde f}_0 \, B(d_\parallel,d_\perp) \, \mu^{-\varepsilon} \over
                \varepsilon} \ .
 \label{mdjzzt}         
\end{eqnarray}
Here, we have employed {\em minimal} subtraction, where only the residues of 
the singular $\varepsilon$ poles were retained; i.e., in the expressions 
$\propto {\widetilde f}_0$, $d$ was replaced with $6 - d_\parallel$, and 
$d_\perp$ with $6 - 2 d_\parallel$ to this order. 
The diffusion constant renormalization is then most conveniently found by 
considering the composite operator $\partial_{q_\perp^2} \, [\lambda q_\perp^2 
+ \Gamma_{{\widetilde S} [{\widetilde S} S]}({\bf q}_\parallel = {\bf 0},
{\bf q}_\perp,\omega = 0)] \vert_{{\bf q}_\perp={\bf 0}}$, with the result
\begin{equation}
        Z^{-1} \, Z_\lambda = 1 + {2 (2-d_\parallel) \over 4-d_\parallel} \, 
        {{\widetilde f}_0 \, B(d_\parallel,d_\perp) \, \mu^{-\varepsilon} \over
                \varepsilon} \ .
 \label{mdjzzl}
\end{equation} 
Equivalently, this combination of $Z$ factors can be establishes by comparing 
the UV singularities in $\partial_{q_\perp^2} \,
\Gamma_{{\widetilde S} S}({\bf q}_\parallel = {\bf 0},{\bf q}_\perp,\omega = 0)
\vert_{{\bf q}_\perp={\bf 0}}$ with the previously established ones in
$\chi({\bf q}_\parallel = {\bf 0},{\bf q}_\perp \to {\bf 0},\omega = 0)$. 
From renormalizing the noise vertex function $\partial_{q_\perp^2} \,
\Gamma_{{\widetilde S}{\widetilde S}}({\bf q}_\parallel = {\bf 0},
{\bf q}_\perp,\omega = 0) \vert_{{\bf q}_\perp={\bf 0}}$, we obtain
\begin{equation}
        Z_S^{-1} \, Z_\lambda = 1 + {{\widetilde f}_0 \, B(d_\parallel,d_\perp)
                \, \mu^{-\varepsilon} \over \varepsilon} \ .
 \label{mdjszl}
\end{equation} 
At last, by means of rather lengthy calculations for the derivatives 
$\partial_{({\bf q}_\perp\cdot{\bf p}_\perp)} \, 
\Gamma_{{\widetilde S}^\alpha S^\beta S^\gamma}(-{\bf q}_\perp,0;
{{\bf q}_\perp \over 2}-{\bf p}_\perp,0;{{\bf q}_\perp \over 2}+{\bf p}_\perp,$
$0) \vert_{{\bf q}_\perp={\bf p}_\perp={\bf 0}}$, $\partial_{q_\perp^2} \, 
\Gamma_{{\widetilde S} S S S}(-{\bf q}_\perp,0;{{\bf q}_\perp \over 3},0;
{{\bf q}_\perp \over 3},0;{{\bf q}_\perp \over 3},0;) 
\vert_{{\bf q}_\perp={\bf 0}}$ we arrive at the coupling constant 
renormalizations 
\begin{eqnarray}
        Z_S \, Z_{\widetilde g} &=& 1 - {d_\parallel \over 3} \, 
        {{\widetilde f}_0 \, B(d_\parallel,d_\perp) \, \mu^{-\varepsilon} \over
                \varepsilon} \ ,
 \label{mdjzsg} \\
        Z_S \, Z_\lambda \, Z_{\widetilde u} &=& 1 - {11 \over 6} \, 
        {{\widetilde u}_0 \, A(d_\parallel,d_\perp) \, \mu^{-\epsilon} \over 
                \epsilon} + \nonumber \\
        &&+ {3-d_\parallel \over 3} \, {{\widetilde f}_0 \, 
        B(d_\parallel,d_\perp) \, \mu^{-\varepsilon} \over \varepsilon} \ .
 \label{mdjzsu}
\end{eqnarray}
For ${\widetilde f}_0 = 0$, Eqs.~(\ref{mdjzzt}) and (\ref{mdjzsu}) reduce to
the one-loop $Z$ factors (\ref{mdbztt}) and (\ref{mdbztu}) for the 
two-temperature model B with $n = 3$, while setting $d_\parallel = 0$ recovers 
the familiar renormalization constants for the equilibrium model J for 
isotropic ferromagnets \cite{modelj,dynfun,erwrev}.

\subsection{Discussion of the RG flow equations}
 \label{mdjflo}

In terms of the renormalized couplings ${\widetilde u}$ and
\begin{equation}
        {\widetilde f} = {{\widetilde g}^2 \over 2 d_\perp \, \lambda^2} \ ,
 \label{jrenmc}
\end{equation}
the one-loop zeta functions for the two-temperature model J become
\begin{eqnarray}
        &&\zeta_S \equiv - \zeta_{\widetilde S} = {d_\parallel (8-d_\parallel) 
                \over 2 (4-d_\parallel)} \, {\widetilde f} \ ,
 \label{mjzets} \\
        &&\zeta = {d_\parallel (6-d_\parallel) \over 2 (4-d_\parallel)} \, 
                {\widetilde f} \ ,
 \label{mjzeta} \\
        &&\zeta_{\widetilde g} = - {\varepsilon \over 2} - {d_\parallel 
        (16-d_\parallel) \over 12 (4-d_\parallel)} \, {\widetilde f} \ ,
 \label{mjzetg} \\
        &&\zeta_\lambda \equiv - \zeta_c = - {d_\parallel^2 - 10 d_\parallel 
                + 8 \over 2 (4-d_\parallel)} \, {\widetilde f} \ ,
 \label{mjzetl} \\
        &&\zeta_\tau = - 2 + {5 \over 6} \, {\widetilde u} 
                - {d_\parallel^2 \over 2 (4-d_\parallel)} \, {\widetilde f} \ ,
 \label{mjzett} \\
        &&\zeta_{\widetilde u} = - \epsilon + {11 \over 6} \, {\widetilde u} 
                - {2 d_\parallel (10 - d_\parallel) \over 3 (4-d_\parallel)} 
                \, {\widetilde f} \ .
 \label{mjzetu}
\end{eqnarray}
Notice that the dynamic coupling constant enters the RG flow for the ``static''
non-linearity ${\widetilde u}$; this is yet another indication that this model
is of genuinely dynamical character.
In the equilibrium limit $d_\parallel = 0$, the statics and dynamics decouple.

The RG beta functions for ${\widetilde u}$ and the effective mode coupling 
(\ref{jrenmc}) of the two-temperature model J read
\begin{eqnarray}
        \beta_{\widetilde u} &=& {\widetilde u} \, \zeta_{\widetilde u} =
        {\widetilde u} \left( - \epsilon + {11 \over 6} \, {\widetilde u}
                - {2 d_\parallel (10 - d_\parallel) \over 3 (4-d_\parallel)} 
                \, {\widetilde f} \right) \ ,
 \label{mjbetu} \\
        \beta_{\widetilde f} &=& 2 {\widetilde f} 
        \left( \zeta_{\widetilde g} - \zeta_\lambda \right) = \nonumber \\
        &&= {\widetilde f} \left( - \varepsilon + 
                {7 d_\parallel^2 - 76 d_\parallel + 48 \over 6 (4-d_\parallel)}
                \, {\widetilde f} \right) \ .
 \label{mjbetf}
\end{eqnarray}
For $d_\parallel = 0$, we thus recover the stable {\em equilibrium} fixed 
points
\begin{equation}
        {\widetilde f}^* = {\varepsilon \over 2} + (\varepsilon^2) \; , \quad
        {\widetilde u}^* = {6 \over 11} \, \epsilon + (\epsilon^2) \ ,
 \label{mjeqfp}
\end{equation}
and critical exponents, see Eqs.~(\ref{mdjstx}) and (\ref{mdjdne}),
\begin{eqnarray}
        \eta &=& 0 + O(\epsilon^2) \; , \quad 
        \nu^{-1} = 2 - {5 \over 11} \, \epsilon + O(\epsilon^2) \ , 
 \label{mjeqst} \\
        z &=& 4 - {\varepsilon \over 2} + O(\epsilon^2) 
                = {d+2 \over 2} + O(\epsilon^2) \ .
 \label{mjeqdy}
\end{eqnarray}
As can be seen here, the correction to $z$ is merely given by the 
$O(\epsilon^2)$ contribution to the static exponent $\eta$ of the 
three-component Heisenberg model.
For, in equilibrium there is an additional identity $Z_g \equiv Z_S$ 
\cite{fnote4}, or $\zeta_g^* \equiv - (\varepsilon + \eta) / 2$.
The condition for the existence of a non-trivial finite fixed point thus
becomes $\zeta_\lambda^* = \zeta_g^* = (d - 6 - \eta) / 2$, or
\begin{equation}
        z \equiv {d + 2 - \eta \over 2} \ .
 \label{mjztex}
\end{equation}

In the full non-equilibrium theory ($d_\parallel > 0$), Eq.~(\ref{mjbetf}) 
still implies that the two-temperature model B fixed point with 
${\widetilde f}^* = 0$ is {\em unstable} for 
$\varepsilon = 6 - d - d_\parallel > 0$.
To one-loop order, the finite positive fixed point
\begin{equation}
        {\widetilde f}^* = {6 (4-d_\parallel) \over 7 d_\parallel^2 
        - 76 d_\parallel + 48} \,  \varepsilon + O(\varepsilon^2,\epsilon^2)
 \label{mjfpft}
\end{equation}
exists only in the interval 
$0 \leq d_\parallel \leq {2 \over 7} (19 - \sqrt{277}) \approx 0.6733$; already
for $d_\parallel = 1$, the RG flow takes the mode coupling to infinity!
(According to the one-loop beta function $\beta_{\widetilde f}$, there is
another regime where ${\widetilde f}^* > 0$, namely 
$4 \leq d_\parallel \leq {2 \over 7} (19 + \sqrt{277}) \approx 10.18$; yet in
such high dimensions $d_c^{\rm st} = 4 - d_\parallel \leq 0$.)
According to Eq.~(\ref{mjbetu}), the divergence of $\beta_{\widetilde f}(\ell)$
under scale transformations as $\ell \to 0$ furthermore drives the ``static'' 
non-linearity to $+ \infty$ as well.
Apparently, the two-temperature model J asymptotically enters a genuine
{\em strong-coupling} regime, which does not allow for a perturbational 
calculation of the critical exponents (for $d_\parallel \geq 0.6733$)
\cite{fnote5}.

Formally though, we may expand about the equilibrium model J, and thus obtain
critical exponents in the limit $d_\parallel \ll 1$.
To first order in $d_\parallel \varepsilon$, we find
\begin{equation}
        {\widetilde f}^* = {\varepsilon \over 2} 
                \left( 1 + {4 \over 3} d_\parallel \right) \; , \quad
        {\widetilde u}^* = {6 \over 11} \, \epsilon
                + {5 \over 11} \, d_\parallel \varepsilon \ ,
 \label{mjexfp}
\end{equation}
leading to the critical exponents
\begin{eqnarray}
        &&\eta = - {3 \over 8} \, d_\parallel \varepsilon \; , \quad
        \eta_S = - {1 \over 2} \, d_\parallel \varepsilon \ ,
 \label{mjexet} \\
        &&\nu^{-1} = 2 - {5 \over 11} \, \epsilon
                - {25 \over 66} \, d_\parallel \varepsilon \ ,
 \label{mjexst} \\
        &&z = 4 - {\varepsilon \over 2} 
                - {1 \over 6} \, d_\parallel \varepsilon \ ,
 \label{mjexdy} \\
        &&\Delta = 1 - {\varepsilon \over 4} 
                - {1 \over 12} \, d_\parallel \varepsilon \ .
 \label{mjexcr}
\end{eqnarray}

Notice, however, that this procedure amounts to an expansion with respect to 
{\em three} dimensional parameters, namely $\epsilon = 4 - d - d_\parallel$,
$\varepsilon = 6 - d - d_\parallel$, and $d_\parallel \varepsilon$.
Moreover, the divergence of the non-expanded fixed point ${\widetilde f}^*$ at
$d_\parallel \approx 0.6733$ indicates that an extrapolation of the formal
results (\ref{mjexst}) and (\ref{mjexdy}) to any physical dimension 
$d_\parallel \geq 1$ is unlikely to work.
On the other hand, we cannot exclude that this divergence is merely a one-loop
artifact, and is cured if one calculates the RG beta functions to higher loop 
orders.
Yet another possibility might well be that the divergence of ${\widetilde f}^*$
and ${\widetilde u}^*$ indicates the absence of a non-equilibrium stationary 
state of the two-temperature model J in the vicinity of its critical point.
A somewhat less drastic implication may be that merely perturbation theory
breaks down, and non-perturbative approaches could possibly characterize the 
scaling behavior at the transition of the two-temperature model J successfully.

At any rate, though, we may draw the following conclusions: 
(i) As opposed to our non-equilibrium version of the SSS model, model J with
dynamical anisotropy remains a genuinely dynamical system, and is very unlikely
to be described by any simple effective equilibrium theory at the critical
point, certainly not by the equilibrium model J. 
(ii) Obviously the reversible mode coupling term in the Langevin equation is
highly relevant, driving the system away from the well-defined two-temperature
model B fixed point towards a strong-coupling regime, which at least to
one-loop order cannot be addressed by means of perturbation theory.

\section{Summary and final remarks}
 \label{sumcon}

In summary, we have extended previous studies on the universality classes of 
non-equilibrium phase transitions by investigating the effect of violating the 
detailed-balance condition in the diffusive dynamics of a conserved field which
is coupled to the order parameter through reversible mode-coupling dynamics.

In a previous work \cite{uwezol}, it was established that the universality 
class of the second-order phase transition in a system with non-conserved order
parameter dynamics is not affected by isotropic breaking of detailed balance. 
Extending this result, we found here that (1) reversible mode coupling 
apparently remains ineffective in generating a new universality class even if 
dynamical anisotropy is present in the diffusive dynamics of the conserved 
field, provided the order parameter itself is {\em non-conserved} and, (2) 
dynamical anisotropy does become relevant if the order parameter itself is
{\em conserved}, in which case reversible mode couplings may drive the
system towards entirely different critical behavior, which apparently cannot be
described by known equilibrium universality classes.

These results give further support to previous observations that the dynamics
of a non-conserved order parameter (with model A being the simplest 
realization) is robust against non-equilibrium perturbations, while the 
dynamics of a conserved order parameter field (model B, if there are no 
reversible mode couplings present) is extremely sensitive to detailed-balance 
breaking through dynamical anisotropy in the system.
One of the surprising features of our results is that dynamical anisotropy,
even in combination with reversible mode coupling terms, cannot destroy the
stability of the equilibrium critical fixed point of the non-equilibrium SSS
model with non-conserved order parameter.
Only in very extreme cases of effectively zero and infinite conserved noise 
temperatures can one find new universality classes, which, however, should well
influence the crossover behavior and the corrections to scaling at the critical
point.
In our non-equilibrium model J, the reversible mode coupling terms had a more
drastic effect, however.
While the two-temperature model B steady state dynamics can be written in the 
form of an effective equilibrium model with elastic or pseudo-dipolar 
long-range interactions, such a simple representation is {\em not} possible for
the two-temperature model J with dynamical anisotropy.
We were not even able to identify a stable and finite renormalization-group 
fixed point (to one-loop order) for this model, but were led to RG runaway 
flows towards a genuinely strong-coupling regime instead (except formally for 
$d_\parallel \ll 1$). 

The fact that the rather complicated combination of relaxation, diffusion, 
reversible mode coupling and dynamical anisotropy does not change the nature of
the second-order phase transitions in systems with {\em non-conserved} order 
parameters, gives us hope that such non-equilibrium phase transitions can be 
understood in terms of a relatively small number of universality classes. 
However, for systems with {\em conserved} order parameters, the situation is 
obviously quite different since both dynamical anisotropy itself, and the 
combination of dynamical anisotropy and reversible mode couplings can lead to 
new universality classes. 
It is clear that the mapping out of the relevant non-equilibrium perturbations 
for the case of conserved order parameter remains an open task.
It should also be noted that most of the studies of this and related problems 
have as yet been restricted to systems with {\em local} currents (heat baths of
different temperatures are attached to the system at every point). 
Problems that are associated with {\em global} currents, such as, e.g., the 
driven lattice gas \cite{bearev}, are much richer and more difficult to 
analyze.
Since the effective interactions observed here carry dipole-like angular 
dependences as well, it may, however, be possible that global currents will not
cause a significant increase in the number of non-equilibrium universality 
classes. 
A systematic investigation of the effects of global currents, involving the 
generalization of the models with local-current into global-current models, 
should clearly be the subject of further studies.

\begin{acknowledgement}

We benefited from discussions with J.~Cardy, E.~Frey, K.~Oerding, 
B.~Schmittmann, F.~Schwabl, S.~Trimper, and R.~Zia.
U.C.T. acknowledges support from the Deutsche Forschungsgemeinschaft through a 
habilitation fellowship, DFG contract no. Ta 177 / 2-1,2.
J.E.S. acknowledges support from the European Commission through a 
TMR Marie Curie fellowship, contract no. ERB-FMBI-CT 97-2816.
Z.R. thanks for partial support by the Hungarian Academy of Sciences 
(Grant OTKA T 019451) and by the EPSRC, United Kingdom (Grant GR / L58088).

\end{acknowledgement}

\appendix
\section{Ward identities for the non-equilibrium SSS model}

In this appendix we derive the basic Ward identities that were used in 
Sec.~\ref{sssren}. In the equilibrium SSS model, one can use the ordinary Ward 
identities between the non-linear susceptibility and the linear susceptibility 
\cite{modele} to reduce the number of $Z$ factors required to renormalize the
theory. However, this procedure is only valid if the response functions 
renormalize multiplicatively, a property which follows from the fact that in 
equilibrium the zero-frequency limit of a multi-linear response function is 
equal to the corresponding static correlation function \cite{sssfth}. Since the
non-equilibrium SSS model does not obey detailed balance, these Ward identities
are not useful in this case. In this appendix we will thus work directly with
the vertex functions, and derive the basic Ward identities which follow from 
the $O(n)$ symmetry of the dynamic functional, and the fact that the 
$M_{0}^{\alpha\beta}$ fields are the generators of this symmetry group. 

We consider the following canonical transformation for the fields
$S_{0}^{\alpha}$ and $M_{0}^{\alpha\beta}$,
\begin{eqnarray}
\delta S_{0}^{\alpha}&=&\epsilon\,\Lambda^{\mu\nu}\{M_{0}^{\mu\nu},
S_{0}^{\alpha}\}=\epsilon\Lambda^{\alpha\nu}S_{0}^{\nu}
\label{apA1}\\
\delta M_{0}^{\alpha\beta}&=&\epsilon\,\Lambda^{\mu\nu}\{M_{0}^{\mu\nu},
M_{0}^{\alpha\beta}\}\nonumber\\
&&=\epsilon(\Lambda^{\alpha\nu}M_{0}^{\nu\beta}
-\Lambda^{\beta\nu}M_{0}^{\nu\alpha}) \ ,
\label{apA2}
\end{eqnarray}
where $\epsilon$ is a small parameter and $\Lambda^{\mu\nu}$ is an arbitrary 
antisymmetric tensor which is constant in space and time. (In this Appendix, we
use Einstein's convention of summation over repeated indices.) This
transformation preserves the Poisson brackets between $S_{0}^{\alpha}$ and 
$M_{0}^{\gamma\delta}$ and between $M_{0}^{\alpha\beta}$ and
$M_{0}^{\gamma\delta}$. If this transformation is supplemented with the 
transformation laws for the auxiliary fields $\tilde{S}_{0}^{\alpha}$ 
and $\tilde{M}_{0}^{\alpha\beta}$, as given by
\begin{eqnarray}
\delta \tilde{S}_{0}^{\alpha}&=&\epsilon\Lambda^{\alpha\nu}
\tilde{S}_{0}^{\nu}
\label{apA3}\\
\delta \tilde{M}_{0}^{\alpha\beta}&=&
\epsilon(\Lambda^{\alpha\nu}\tilde{M}_{0}^{\nu\beta}
-\Lambda^{\beta\nu}\tilde{M}_{0}^{\nu\alpha}) \ ,
\label{apA4}
\end{eqnarray}
then one can show that the Janssen-de Dominicis functional 
$J[\{ {\widetilde S}_0^\alpha \}, \{ S_0^\alpha \},
        \{ {\widetilde M}_0^{\alpha \beta} \}, \{ M_0^{\alpha \beta} \}]$
is invariant with respect to the joint transformation (\ref{apA1}) to
(\ref{apA4}). 

If the tensor $\Lambda^{\mu\nu}(t)$ is now allowed to depend on time,
then the dynamic functional is no longer invariant under the transformations 
(\ref{apA1}) to (\ref{apA4}), but picks up the extra terms
\begin{eqnarray}
\delta J&=&-\epsilon\int\,d^{d}x\,\int\,dt\,\dot{\Lambda}^{\alpha\beta}(t)\,
[\tilde{S}S]^{\alpha\beta}({\bf x},t)-\nonumber\\
&&-\epsilon\int\,d^{d}x\,\int\,dt\,\dot{\Lambda}^{\alpha\beta}(t)\,
[\tilde{M}M]^{\alpha\beta}({\bf x},t) \ ,
\label{apA5}
\end{eqnarray}
where $[\tilde{S}S]_{0}^{\alpha\beta}({\bf x},t)$ and
$[\tilde{M}M]_{0}^{\alpha\beta}({\bf x},t)$ are composite operators which are
defined as 
\begin{eqnarray}
[\tilde{S}S]^{\alpha\beta}({\bf x},t)&=&
\frac{1}{2}\,[\,
\tilde{S}_{0}^{\alpha}({\bf x},t)
S_{0}^{\beta}({\bf x},t)-\nonumber\\
&&\mbox{}-\tilde{S}_{0}^{\beta}({\bf x},t)
S_{0}^{\alpha}({\bf x},t)\,] \ ,
\label{apA6}
\end{eqnarray}
and
\begin{eqnarray}
[\tilde{M}M]^{\alpha\beta}({\bf x},t)&=&
\frac{1}{2}\,[\,\tilde{M}_{0}^{\alpha\nu}({\bf x},t)
M^{\beta\nu}({\bf x},t)-\nonumber\\
&&\mbox{}-\tilde{M}_{0}^{\beta\nu}({\bf x},t)
M_{0}^{\alpha\nu}({\bf x},t)\,] \ ,
\label{apA7}
\end{eqnarray}
and the ``$\cdot$'' stands for differentiation with respect to time. In order 
to proceed, we consider the generating functional
\begin{eqnarray}
&&Z[h^{\alpha},\tilde{h}^{\alpha},H^{\alpha\beta},\tilde{H}^{\alpha\beta},
{\cal J}^{\alpha\beta},J^{\alpha\beta},{\cal L}^{\alpha}]=\nonumber\\
&&=\int{\cal D}[\{ S_0^\alpha \}] {\cal D}[\{ i {\widetilde S}_0^
\alpha \}] \int {\cal D}[\{M_0^{\alpha \beta} \}]
        {\cal D}[\{ i {\widetilde M}_0^{\alpha \beta} \}] \times \nonumber \\
        &&\times \, e^{J[\{ {\widetilde S}_0^\alpha \}, \{ S_0^\alpha \},
        \{ {\widetilde M}_0^{\alpha \beta} \}, \{ M_0^{\alpha \beta} \}]}
\nonumber\\
&&\mbox{}\times e^{\int_{{\bf x},t}h^{\alpha}S_{0}^{\alpha}+
\tilde{h}^{\alpha}\tilde{S}_{0}^{\alpha}+
\frac{1}{2}(H^{\alpha\beta}M_{0}^{\alpha\beta}+
\tilde{H}^{\alpha\beta}\tilde{M}_{0}^{\alpha\beta})}\nonumber\\
&&\mbox{}\times e^{\frac{1}{2}\int_{{\bf x},t}
({\cal J}^{\alpha\beta}[\tilde{S}S]_0^{\alpha\beta}+
 J^{\alpha\beta}[\tilde{M}M]_0^{\alpha\beta}
+{\cal L}^{\alpha}[\tilde{M}S]_0^{\alpha})} \ ,
\label{apA8}
\end{eqnarray}
where we have introduced a source term for the fields 
$S_{0}^{\alpha}$, $\tilde{S}_{0}^{\alpha}$, $M_{0}^{\alpha\beta}$,
and $\tilde{M}_{0}^{\alpha\beta}$, $[\tilde{S}S]_{0}^{\alpha\beta}$ and 
$[\tilde{M}M]_{0}^{\alpha\beta}$ in $Z$, and also a source term for the
composite field $[\tilde{M} S]_{0}^{\alpha}=\tilde{M}^{\alpha\nu}S_{0}^{\nu}$,
since this composite operator also enters in the definition of the linear
susceptibility (\ref{opvsus}). One can now obtain the transformation laws for
any composite operator from the transformations (\ref{apA1}) to (\ref{apA4}).

If one now applies the transformation (\ref{apA1}) to (\ref{apA4}) with a 
time-dependent parameter $\Lambda^{\mu\nu}(t)$ to the dynamic functional $Z$, 
then it is easy to see that the two terms generated by the transformation, as
given in Eq.~(\ref{apA5}), can be absorbed in the transformation law for the
sources of $[\tilde{S}S]_{0}^{\alpha\beta}$ and 
$[\tilde{M}M]_{0}^{\alpha\beta}$.

One ends up with the following identity
\begin{eqnarray}
&&Z[h^{\alpha},\tilde{h}^{\alpha},H^{\alpha\beta},\tilde{H}^{\alpha\beta},
{\cal J}^{\alpha\beta},J^{\alpha\beta},{\cal L}^{\alpha}]=\nonumber\\
&&Z[h^{\alpha}-\epsilon\Lambda^{\alpha\beta}h^{\beta},
\tilde{h}^{\alpha}-\epsilon\Lambda^{\alpha\beta}\tilde{h}^{\beta},\nonumber\\
&&\:\;\;
H^{\alpha\beta}-\epsilon(\Lambda^{\alpha\nu}H^{\nu\beta}-\Lambda^{\beta\nu}
H^{\nu\alpha}),\nonumber\\
&&\;\;\;\tilde{H}^{\alpha\beta}
-\epsilon(\Lambda^{\alpha\nu}\tilde{H}^{\nu\beta}-\Lambda^{\beta\nu}
\tilde{H}^{\nu\alpha}),\nonumber\\
&&\;\;\;{\cal J}^{\alpha\beta}-
\epsilon(\Lambda^{\alpha\nu}{\cal J}^{\nu\beta}-\Lambda^{\beta\nu}
{\cal J}^{\nu\alpha})-2\epsilon\dot{\Lambda}^{\alpha\beta},\nonumber\\
&&\;\;\;J^{\alpha\beta}-
\epsilon(\Lambda^{\alpha\nu}J^{\nu\beta}-\Lambda^{\beta\nu}
 J^{\nu\alpha})-2\epsilon\dot{\Lambda}^{\alpha\beta},\nonumber\\
&&\;\;\;{\cal L}^{\alpha}-\epsilon\Lambda^{\alpha\beta}{\cal L}^{\beta}]\, \ .
\label{apA9}
\end{eqnarray}

Expanding this identity to first order in $\epsilon$, and using the fact that 
$\Lambda^{\alpha\beta}(t)$ is antisymmetric but otherwise arbitrary, one 
obtains the following relation for the vertex functions
\begin{eqnarray}
&&\int\,d^{d}x\left\{\left(s^{\alpha}_{0}\frac{\delta\Gamma}{\delta 
s_{0}^{\beta}}
-s^{\beta}_{0}\frac{\delta\Gamma}{
\delta s_{0}^{\alpha}}\right)+\right.\nonumber\\
&&\mbox{}+\left(\tilde{s}^{\alpha}_{0}\frac{\delta\Gamma}{\delta 
\tilde{s}_{0}^{\beta}}
-\tilde{s}^{\beta}_{0}\frac{\delta\Gamma}{
\delta \tilde{s}_{0}^{\alpha}}\right)\nonumber\\
&&\mbox{}-\left(m^{\alpha\nu}_{0}\frac{\delta\Gamma}{\delta 
m_{0}^{\nu\beta}}
-m^{\beta\nu}_{0}\frac{\delta\Gamma}{
\delta m_{0}^{\nu\alpha}}\right)\nonumber\\
&&\mbox{}-\left(\tilde{m}^{\alpha\nu}_{0}\frac{\delta\Gamma}{\delta 
\tilde{m}_{0}^{\nu\beta}}
-\tilde{m}^{\beta\nu}_{0}\frac{\delta\Gamma}{
\delta \tilde{m}_{0}^{\nu\alpha}}\right)\nonumber\\
&&\mbox{}-\left({\cal J}^{\alpha\nu}_{0}\frac{\delta\Gamma}{\delta 
{\cal J}_{0}^{\nu\beta}}
-{\cal J}^{\beta\nu}_{0}\frac{\delta\Gamma}{
\delta {\cal J}_{0}^{\nu\alpha}}\right)\nonumber\\
&&\mbox{}-\left(J^{\alpha\nu}_{0}\frac{\delta\Gamma}{\delta 
J_{0}^{\nu\beta}}
-J^{\beta\nu}_{0}\frac{\delta\Gamma}{
\delta J_{0}^{\nu\alpha}}\right)\nonumber\\
\nonumber\\
&&\mbox{}+\left({\cal L}^{\alpha}_{0}\frac{\delta\Gamma}{\delta 
{\cal L}_{0}^{\beta}}
-{\cal L}^{\beta}_{0}\frac{\delta\Gamma}{
\delta {\cal L}_{0}^{\alpha}}\right)\nonumber\\
&&\left.+2\frac{\partial}{\partial t}\left(\frac{\delta\Gamma}{
\delta {\cal J}_{0}^{\alpha\beta}}+\frac{\delta\Gamma}{
\delta J_{0}^{\alpha\beta}}\right)\right\}=0\, \ ,
\label{apA10}
\end{eqnarray}
where $s_{0}^{\alpha}({\bf x},t) = \langle S_{0}^{\alpha}({\bf x},t)\rangle$, 
$\tilde{s}_{0}^{\alpha}({\bf x},t) = \langle \tilde{S}_{0}^{\alpha}({\bf x},t)
\rangle$, and $m_{0}^{\alpha\beta}({\bf x},t) = 
\langle M_{0}^{\alpha\beta}({\bf x},t)\rangle$, 
$\tilde{m}_{0}^{\alpha\beta}({\bf x},t) = 
\langle \tilde{M}_{0}^{\alpha\beta}({\bf x},t)\rangle$.

A similar identity can be obtained for the correlation functions. One can then 
derive the usual Ward identities for multi-linear response functions from this 
identity. This establishes the equivalence of the two procedures in the 
equilibrium case.

Taking the variational derivative of (\ref{apA10}) with respect to 
$\tilde{m}^{\gamma\delta}({\bf x},t)$ and $m^{\nu\zeta}({\bf x},t)$, and 
setting the source terms to zero, one obtains, after taking the Fourier 
transform, the identity
\begin{eqnarray}
&&-i\tilde{\omega}
\left\{\Gamma_{0\tilde{M}M[\tilde{S}S]}^{\gamma\delta\eta\zeta
\alpha\beta}({\bf q},\omega\,;{\bf 0},\tilde{\omega})+
\Gamma_{0\tilde{M}M[\tilde{M}M]}^{\gamma\delta\eta\zeta
\alpha\beta}({\bf q},\omega\, ;{\bf 0},\tilde{\omega})\right\}=\nonumber\\
&&= {1 \over 2} [\,\delta^{\alpha\eta}(\delta^{\beta\delta}\delta^{\gamma\zeta}
-\delta^{\beta\gamma}\delta^{\delta\zeta})+
\delta^{\alpha\zeta}(\delta^{\beta\gamma}\delta^{\delta\eta}
-\delta^{\beta\delta}\delta^{\gamma\eta})\nonumber\\
&&\mbox{}+\delta^{\alpha\gamma}(\delta^{\beta\eta}\delta^{\delta\zeta}
-\delta^{\beta\zeta}\delta^{\delta\eta})
\mbox{}+\delta^{\alpha\delta}(\delta^{\beta\zeta}\delta^{\gamma\eta}
-\delta^{\beta\eta}\delta^{\gamma\zeta})\,]\nonumber\\
&&\times\{\Gamma_{0\tilde{M}M}({\bf q},\omega+\tilde{\omega})-
\Gamma_{0\tilde{M}M}({\bf q},\omega)\} \ ,
\label{apA11}
\end{eqnarray}
where we have used the tensor properties of 
$\Gamma_{0\tilde{M}M}^{\alpha\beta\gamma\delta}$. This identity relates the 
vertex functions with one insertion of the composite operators 
$[\tilde{S}S]^{\alpha\beta}$ and $[\tilde{M}M]^{\alpha\beta}$ to vertex 
functions with no insertions. Taking the variational derivatives of 
Eq.~(\ref{apA10}) with respect to the other fields, one can obtain similar Ward
identities for other vertex functions.

These Ward identities show that no multiplicative re\-nor\-ma\-lization is
needed for the composite ope\-ra\-tors $ [\tilde{S}S]^{\alpha\beta}$ and 
$[\tilde{M}M]^{\alpha\beta}$, i.e. $Z_{[\tilde{S}S]}=Z_{[\tilde{M}M]}=1$.
However, these identities do not exclude the need of
additive renormalization for these operators provided that this 
renormalization is $\propto q^2$ \cite{dynfun}.

We now consider Eq.~(\ref{opvsus}), which follows from the identity
\begin{equation}
\left\langle\frac{\delta J}{\delta M^{\alpha\beta}_0({\bf x},t)}
\right\rangle\,+\,H^{\alpha\beta}({\bf x},t)=0 \ ,
\label{apA12}
\end{equation}
which in turn can be proven using the fact that the path integral of a 
functional derivative vanishes. If one now takes the functional derivative of
this equation with respect to $\tilde{m}^{\gamma\delta}({\bf x}',t')$ and
performs a Fourier transformation, one obtains Eq.~(\ref{opvsus}).

The renormalized version of (\ref{opvsus}) can be obtained if one uses the
identity $Z_{[\tilde{S}S]}=1$ and the definition of the renormalized diffusion 
constant $\partial_{q^2}\Gamma_{\tilde{M}M}({\bf q}\, \omega)\mid_{NP}=D$ in 
Eq.~(\ref{opvsus}). We thus find
\begin{eqnarray}
&&\Gamma_{\tilde{M}M}({\bf q},\omega)
=i\omega+Dq^2+\nonumber\\
&&\mbox{}+2g\mu^{\epsilon/2}A_{d}^{-1/2}
(Z_{M}Z_{g})^{-1/2}\Gamma_{\tilde{M}[\tilde{S}S]}({\bf q},\omega) \, ,
\label{apA13}
\end{eqnarray}
where the renormalized vertex function 
$\Gamma_{\tilde{M}[\tilde{S}S]}({\bf q},\omega)$ is defined by
\begin{eqnarray}
&&\Gamma_{\tilde{M}[\tilde{S}S]}({\bf q},\omega)\equiv Z_{\tilde{M}}^{-1/2}
\label{apA14} \\
&&\mbox{}\times(
\Gamma_{0 \tilde{M}[\tilde{S}S]}({\bf q},\omega)-q^{2}\partial_{q^2}
\Gamma_{0 \tilde{M}[\tilde{S}S]}({\bf q},\omega)\mid_{NP}) \, , \nonumber
\end{eqnarray}
and where the second term on the right-hand-side follows from the definition of
the renormalized diffusion constant and corresponds to an additive 
renormalization needed to render $[\tilde{S}S]^{\alpha\beta}$ finite.

Hence, since  all the quantities in (\ref{apA13}) are finite, we conclude that
\begin{equation}
Z_{g}Z_{M}=1
\label{apA15}
\end{equation}
must hold to {\it all orders} in perturbation theory. If one takes the ratio of
Eqs.~(\ref{zsmrel}) and (\ref{zstrel}), one can check explicitly that this 
identity holds to one loop-order.

A similar set of Ward identities can be derived for the equilibrium model J
\cite{dynfun,fnote4}, but it is unclear if these identities still hold in the
non-equilibrium {\em effective} model J with long-range elastic forces 
considered in Sec.~\ref{mdjren}.


\begin{thebibliography}{}
\bibitem{fnote2} Notice that we have employed a different convention for the
                renormalization constants as the one chosen in 
                Ref.~\cite{uwezol}. 
                Of course, these different definitions have no influence on the
                RG flow functions, which describe the physical scaling 
                behavior, see Sec.~\ref{sssflo}.
\bibitem{fnote4} In equilibrium, though, there exists a relation between linear
                and non-linear response functions \cite{modele,sssfth}; for the
                SSS model, see Eq.~(3.14) in Ref.~\cite{uwezol}.
                For the equilibrium model J, one may readily derive a similar
                connection between the linear and non-linear order parameter
                susceptibility, which leads to $Z_g \equiv Z_S$ 
                \cite{modelj,dynfun}.
                However, in the non-equilibrium generalization of model J, the
                non-linear response function acquires novel UV singularities; 
                therefore $Z_g$ remains undetermined through other 
                renormalization constants, and must be computed separately.
\end{thebibliography}
\end{document}